\documentclass[reprint,amsmath,amssymb,aps,prb,longbibliography,floatfix]{revtex4-2}
\usepackage[T1]{fontenc}
\usepackage[utf8]{inputenc}
\usepackage{textcomp}
\usepackage{graphicx}
\usepackage{bm}
\usepackage[colorlinks,bookmarks=true,citecolor=blue,linkcolor=red,urlcolor=blue]{hyperref}
\newcommand{\elliptic}[5][\scriptstyle]{\vartheta\left[\begin{array}{c}{{#1 #2}}\\{#1 #3}\end{array}\right]\left(#4\middle|#5\right)}
\providecommand{\theHequation}{\arabic{equation}}
\providecommand{\theHfigure}{\arabic{figure}}
\providecommand{\theHtable}{\arabic{table}}
\providecommand{\theHsection}{\arabic{section}}
\providecommand{\theHsubsection}{\theHsection.\arabic{subsection}}
\begin{document}

\title{Composite fermions in ideal Chern bands} 

\author{Songyang Pu}
\affiliation{Department of Physics, Washington University in St. Louis, St. Louis, Missouri 63130, USA}
\author{Liangtao Peng}
\affiliation{Department of Physics, Washington University in St. Louis, St. Louis, Missouri 63130, USA}
\author{Shaffique Adam}
\affiliation{Department of Physics, Washington University in St. Louis, St. Louis, Missouri 63130, USA}

\date{October 1, 2026}

\begin{abstract}
The composite-fermion framework maps the strongly correlated fractional quantum Hall state onto a weakly interacting integer quantum Hall state in a reduced effective magnetic field. We ask how this framework is modified for fractional Chern insulators, where the role of a uniform magnetic field is replaced by the nonuniform Berry curvature of a Chern band. Specifically, we construct composite-fermion wave functions for Jain states and their quasiparticles and quasiholes in Aharonov--Casher bands, a widely used model for ideal Chern bands. We find that these states overlap closely with exact eigenstates over a broad range of field modulation. Using these wave functions, we demonstrate that the emergent composite-fermion band structure goes beyond the conventional picture of effective Landau levels, correctly capturing the quasiparticle dispersion. Remarkably, we find that an isolated quasiparticle band carries a total Berry flux that is different from a Landau level. These results establish a composite-fermion description for mobile fractional Chern insulator anyons and provide a framework to understand their stability.
\end{abstract}

\maketitle

{\it Introduction.---}  A central strategy for understanding strongly correlated matter is to identify the appropriate emergent quasiparticles. Once the interacting electrons are expressed in terms of these quasiparticles, the residual interactions are weak, and the problem becomes tractable.  A paradigmatic example is the composite fermion (CF) framework where new quasiparticles are formed by binding an electron to an even number of vortices ($2p$)~\cite{Jain89,Jain07}. Composite-fermion theory then maps the strongly correlated electrons at fillings $\nu=n/(2pn\pm1)$ to the weakly interacting CFs occupying $n$ effective Landau levels, 
called $\Lambda$ levels. For $n=1$, the CF wave function reduces to Laughlin's wave function~\cite{PhysRevLett.50.1395}, so the CF construction can be viewed as its generalization to the Jain sequences of Abelian fractional quantum Hall states.

The CF framework provides accurate many-body wave functions for ground states and excitations across the Jain sequence with no adjustable parameters.  Moreover, CF wavefunctions can be evaluated for system sizes well beyond the reach of exact diagonalization.  The framework explains the sequences of prominent fractions observed experimentally~\cite{Du1993}, predicts the composite fermion Fermi sea at $\nu=1/2$~\cite{Halperin1993Theory,Willett1993Experimental,Kang1993How}, gives the dispersion of neutral CF excitons observed in inelastic light scattering~\cite{Kang2001Observation} and microwave absorption~\cite{Kukushkin09} experiments, and explains the observed Bloch ferromagnetism phenomenon \cite{Hossain2021bloch}. The emergent $\Lambda$ levels that govern the internal structure of individual CF quasiparticles and quasiholes~\cite{Jeon_2003} give spectroscopic signatures~\cite{Pu2024Fingerprints,Gattu2024Scanning} that have been recently observed in local tunneling measurements~\cite{Hu_2025}.

Fractional Chern insulators (FCIs) extend fractional quantum Hall physics to topological bands without an external magnetic field \cite{Tang2011High,Sun2011Nearly,Neupert2011Fractional,Sheng2011Fractional}. Their recent observation in twisted bilayer MoTe$_2$ \cite{Zeng2023thermodynamic,Cai2023Signatures,Park2023Observation,Xu2023Observation} and rhombohedral multilayer graphene aligned with hBN \cite{Lu2024Fractional} raises a fundamental question: to what extent is the CF framework applicable to fractional Chern insulators and how is the CF description modified by the underlying crystal?  One key difference is that a continuous magnetic-translation symmetry guarantees that the energy of an elementary charged excitation is independent of its guiding-center position in a Landau level,  making the charged quasiparticles dispersionless and macroscopically degenerate. 
On the other hand, in an FCI, the symmetry is reduced to discrete lattice translations, allowing anyons to acquire a 
dispersion even when the electronic Chern band is exactly flat~\cite{Liu2025Characterization,Gon_alves_2026,Schleith2026Anyon}.  This has been recently demonstrated for Laughlin quasiholes in Aharonov--Casher bands~\cite{yan2025anyon,iyer2026dispersion}, a widely used and analytically controlled realization of ideal Chern bands ~\cite{Ledwith2020Fractional,Wang2021Exact,Shi2024Adiabatic,SM}.  

The experimentally observed FCIs extend beyond the Laughlin states and predominantly follow the Jain sequences. This raises the question of how to incorporate the anyon dispersion with the $\Lambda$ level structure. Do the CFs develop a dispersive band structure beyond the flat $\Lambda$ level, or does the dispersion emerge from the residual interaction between CFs? Answering this question requires applying the CF theory to FCI. Hamiltonian and projective approaches have generalized CF theory to Chern bands~\cite{Murthy2003Hamiltonian,Murthy2012Ham,Hu2024Hyper,Hu2026composite,Jin2025,Lin2026hyper}, and other constructions map continuum quantum Hall trial states into Chern bands~\cite{Qi2011Generic,Wu2012Bloch,Ledwith2020Fractional,Wang2021Exact}. Yet a controlled calculation of the dispersion of individual CF quasiparticles and quasiholes across the Jain sequence is still lacking.

In this Letter, we extend the composite-fermion framework to construct wave functions for Jain ground states and their quasiparticle (QP) and quasihole (QH) excitations in $|\mathcal{C}|=1$ ideal Chern bands. We find that charged CF excitations form dispersive Bloch bands even though the underlying electronic band is exactly flat, demonstrating that the anyon dispersion mainly originates from the band structure of a single composite fermion, with negligible effect from the residual interaction between composite fermions. More strikingly, the isolated quasiparticle band carries a different Chern number compared to a single-electron Landau level. For the Jain state at filling $\nu=n/(2pn+1)$, its Chern number over the electronic Brillouin zone is equivalent to the ground state degeneracy
\begin{equation}
\mathcal{C}_{\rm QP}\equiv \frac{1}{2\pi}\sum_{{\bf k}\in\rm BZ}\Omega_{\rm QP}(\mathbf{k})=2pn+1.
\label{CQP}
\end{equation}
Thus, the emergent quasiparticle band retains a direct and quantized imprint of its parent Jain state. This dispersive and topological band structure determined by the CF framework will control the properties of proposed exotic phases including anyon molecules, compressible anyon fluids, and anyon superconductors \cite{Gattu2025Molecular,Wang2026anyon,Xu2025dynamics,Laughlin1988Super,Laughlin1988The,Nosov_2026,han2026anyon,Pichler_2026}. Moreover, as long as the QP band remains isolated, the exactly quantized Chern number provides a complementary diagnostic of the FCI phase even when the finite-size ground-state degeneracy is imperfect, as tested for a microscopic model for twisted bilayer MoTe$_2$ shown in the SM~\cite{SM}.

\begin{figure}[]
    \includegraphics[width=0.98\linewidth]{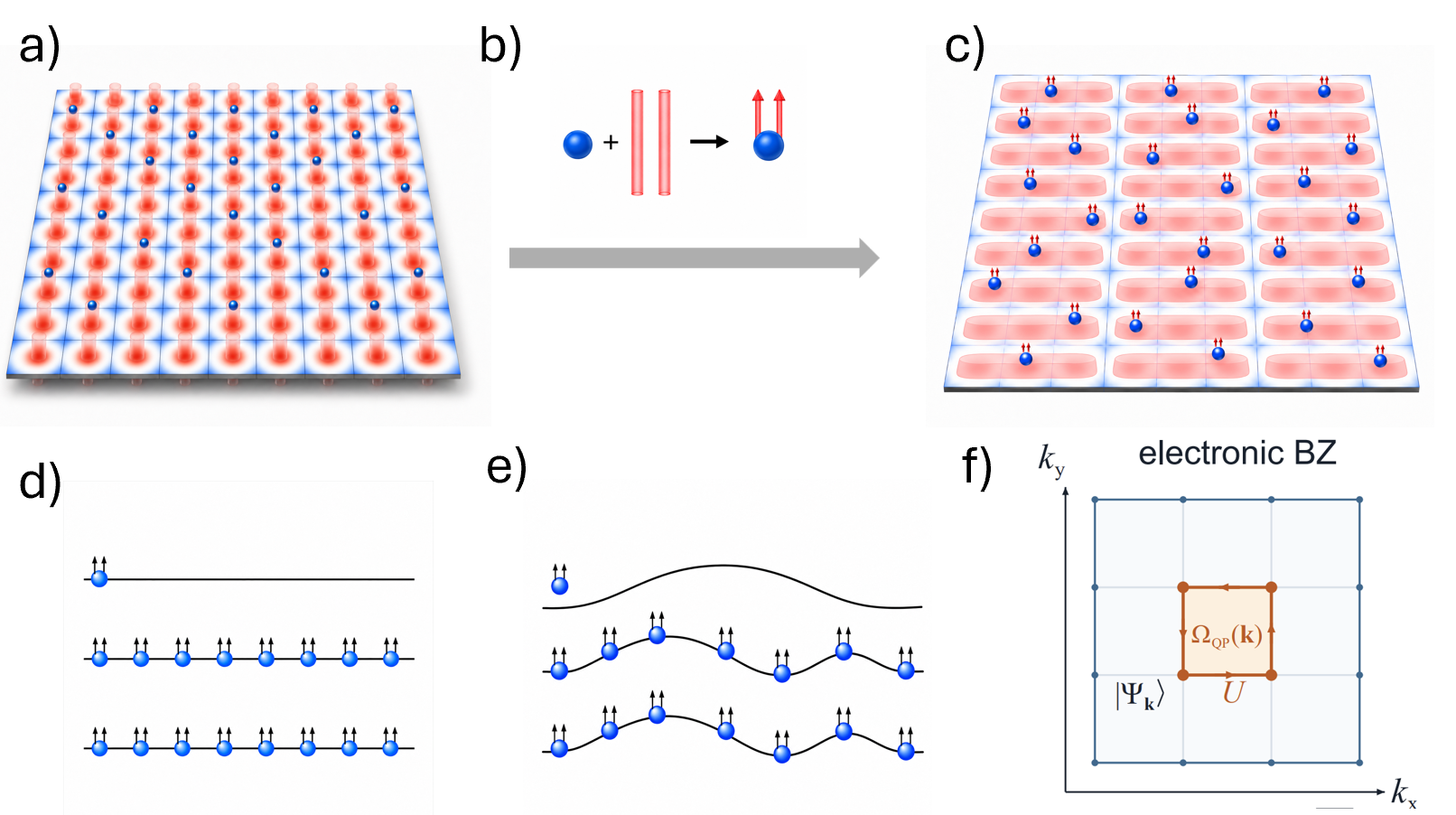}
    \caption{Illustration of our model. (a) Model system in which each unit cell contains one unit of magnetic flux, distributed unevenly within the cell. (b) Each electron is combined with two magnetic vortices to form a composite fermion. (c) At one-third filling, this attachment leaves one unit of magnetic flux for every three original unit cells, so the effective lattice becomes three times larger in one direction. (d) In an ordinary magnetic field, the allowed states of these quasiparticles have the same energy, giving flat quasiparticle bands. (e) In a crystal that reproduces the same quantum behavior without an external magnetic field, these states instead acquire different energies and form dispersing bands, even though the original electron band is perfectly flat. (f) Definition of $C_{\mathrm{QP}}$. Vertices represent many-body QP momentum states, and arrows denote normalized density links $U$. The counterclockwise loop defines the plaquette Berry flux $\Omega_{\rm QP}({\bf  k})=-\operatorname{Arg}\prod U$. Summing over the electronic Brillouin zone gives $\mathcal{C}_{\mathrm{QP}}=\sum_{\bf   k}\Omega_{\rm QP}({\bf k})/(2\pi)$.}
    \label{fig0}
\end{figure}

{\it Model.---} We consider ideal $|\mathcal C|=1$ bands whose Bloch orbitals admit the
factorization $\varphi_{\boldsymbol\kappa}(\mathbf r)
=\chi(\mathbf r)\psi^{\mathrm{AC}}_{\boldsymbol\kappa}(\mathbf r)$,
where $\chi$ is momentum independent and satisfies
$\chi^\dagger(\mathbf r)\chi(\mathbf r)=1$
pointwise~\cite{Guerci2025layer,Shi2024Adiabatic,Estienne2023ideal}.
For density--density interactions that couple identically to all
internal components, $\chi$ cancels from the projected interaction
matrix elements. We therefore work with the scalar Aharonov--Casher
(AC) Hamiltonian~\cite{Aharonov1979Ground,SM},
\begin{equation}
 H_{\mathrm{AC}}=\frac{\Pi^\dagger\Pi}{2m},\qquad
 \Pi=\Pi_x+i\Pi_y,\qquad \Pi_\mu=-i\partial_\mu+A_\mu,
 \label{eq:model-ac-hamiltonian}
\end{equation}
in units $e=\hbar=1$. We choose
$\mathbf B=\boldsymbol\nabla\times\mathbf A=-B(\mathbf r)\hat{\mathbf z}$,
with
\begin{equation}
 B(\mathbf r)=B_0\!\left[
 1+\gamma\!\left(\sin\frac{2\pi x}{a_x}
                  +\sin\frac{2\pi y}{a_y}\right)\right],
 \label{eq:model-field}
\end{equation}
and one flux quantum per unit cell, $B_0a_xa_y=2\pi$. As illustrated in Fig.~\ref{fig0}, we place an $N_x\times N_y$ array of unit cells on a torus with $N_\phi=N_xN_y$ flux quanta, with the field varying within each cell for $\gamma\ne0$. The parameter $\gamma$ controls the magnetic-field modulation. The zero modes $\Pi\psi=0$ reduce to the lowest Landau level at
$\gamma=0$ and remain exactly flat for nonzero $\gamma$.
We project the bare Coulomb interaction into this band and resolve
the many-body spectrum by total crystal momentum. The torus geometry,
explicit orbitals, and momentum labels are given in the End Matter.

{\it Composite-fermion construction.---} In conventional composite-fermion (CF) theory, attaching $2p$ vortices to each electron reduces the number of flux quanta seen by the CFs to $N_\phi^*=N_\phi-2p N_e$, as illustrated in Fig.~\ref{fig0}(b-c). At the Jain filling
$\nu=n/(2pn+1)$ on the torus, one has $N_\phi^*=N_e/n$, and the CFs form an integer quantum Hall state with $n$ filled effective Landau levels, or $\Lambda$ levels.

In the AC model, however, the physical magnetic field is modulated over an $N_x\times N_y$ lattice. Given that $N_\phi^*$ might not be compatible with this lattice decomposition, the effective CF orbitals do not naturally transform as Bloch states under the primitive magnetic translations of the physical AC lattice. We resolve this mismatch in two steps. First, we construct a CF state directly within the AC band without imposing a definite lattice momentum. Second, we project this state onto an irreducible momentum sector of the physical lattice magnetic-translation group.

The CF wave function at $\nu={n\over 2pn+1}$ can be written as
\begin{widetext}
\begin{equation}
    \Psi_{\rm CF}=\mathcal{P}_{\mathrm{AC}}e^{\sum_i Q_i}
        \begin{vmatrix}
        f^{(0)}_0(z_1)\cdots f^{(0)}_0(z_{N_e})\\
        \vdots\\
        f^{(0)}_{N_\phi^*-1}(z_1)\cdots f^{(0)}_{N_\phi^*-1}(z_{N_e})\\
        f^{(1)}_0(z_1)\cdots f^{(1)}_0(z_{N_e})\\
        \vdots\\
        f^{(n-1)}_{N_\phi^*-1}(z_1)\cdots f^{(n-1)}_{N_\phi^*-1}(z_{N_e})\\
    \end{vmatrix}\Bigg [\vartheta \begin{bmatrix}
(N_e-1)/2 \\ (N_e-1)/2
\end{bmatrix}\Bigg({\frac{Z}{L_x}} \Bigg | \tau \Bigg) \prod_{i<j}  \vartheta 
\begin{bmatrix}
\frac{1}{2} \\ \frac{1}{2}
\end{bmatrix}
\Bigg( \frac{z_i-z_j}{L_x}\Bigg | \tau \Bigg ) \Bigg ]^{2p},
\end{equation}
\end{widetext}
\noindent where the gauge factor $Q_i$ is given by Eq.~\ref{eq:model-Q}, $f^{(m)}_n(z_j)$ is the $\Lambda$ level orbital given in the SM \cite{Pu17,SM} with the special case $m=0$ given in Eq.~\ref{eq:model-theta-orbitals}.

The operator $\mathcal{P}_{\mathrm{AC}}$ projects the state into the AC band. We employ a generalized Jain--Kamilla-type prescription to construct CF trial states within the AC band \cite{Jain1997,Pu17}. The resulting wave functions satisfy the zero-mode condition $\Pi_i\Psi_{\mathrm{CF}}=0$ for every particle $i$. Here ``projection'' denotes this construction within the AC band and does not, in general, coincide with orthogonal projection of the original unprojected CF ansatz. The explicit form is provided in the Supplemental Material \cite{SM}. At $\gamma=0$, this construction reduces to the conventional torus CF wave function in the lowest Landau level.

As mentioned above, the CF wave functions are not the eigenstates of center-of-mass magnetic translation $t_{\rm CM}(a_{x(y)})=\prod_{i=1}^{N_e}t_i(a_{x(y)})$. Meanwhile, we can apply a momentum projector to project it to $(K_x,K_y)$ sector up to an overall normalization factor 
\begin{align}
    \mathcal{P}_{(K_x,K_y)}=&\sum_{m_1=0}^{N_x-1}\sum_{m_2=0}^{N_y-1}e^{-i2\pi\left({K_xm_1\over N_x}+{m_1\phi_1N_e\over2\pi N_x}+{K_ym_2\over N_y}+{m_2\phi_2N_e\over2\pi N_y}\right)} \nonumber \\
    &t_{\rm CM}^{m_1}(a_x)t_{\rm CM}^{m_2}(a_y)
    \label{PK1}
\end{align}
such that $t_{\rm CM}(a_\mu)\mathcal{P}_{(K_x,K_y)}\Psi_{\rm CF}=e^{i(2\pi K_\mu+N_e\phi_\mu)/N_\mu}\mathcal{P}_{(K_x,K_y)}\Psi_{\rm CF}$. Since $[t_i(a_\mu),\Pi_i]=0$, the momentum projection does not alter the condition
$\Pi_i|\Psi_{\mathrm{CF}}\rangle=0$.

\begin{figure}[]
    \includegraphics[width=0.98\linewidth]{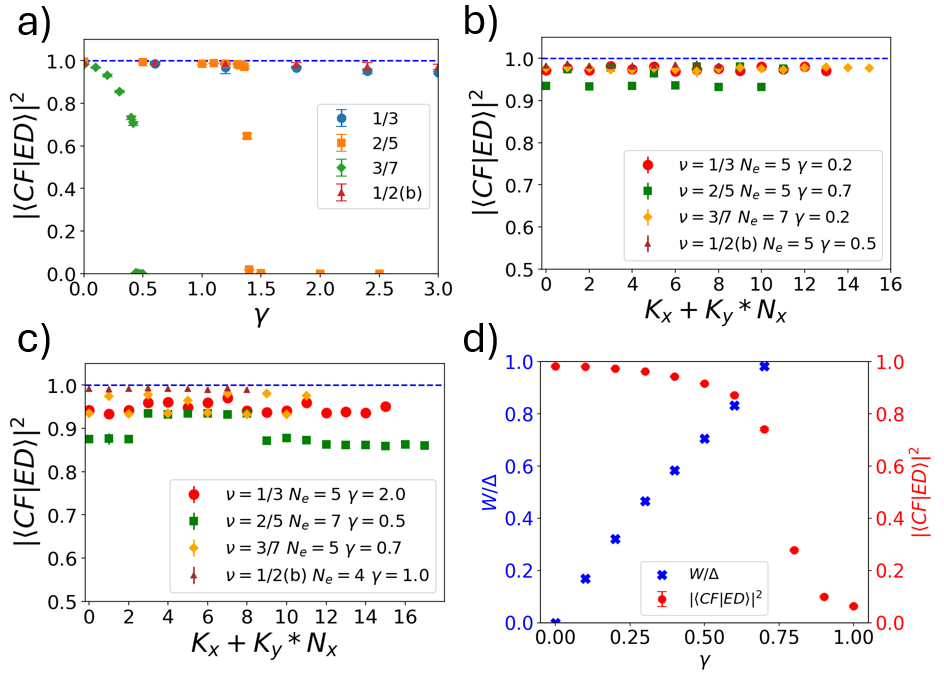}
    \caption{(a) The squared overlaps between the exact ground states and composite fermion ground states. We choose $N_e=5$ for $\nu=1/3,1/2$ and $N_e=6$ for $\nu=2/5, 3/7$  (We only choose one momentum sector for each filling.) The $\nu=1/3,2/5,3/7$ are fermionic Jain states and $\nu=1/2$ is a bosonic Laughlin (Jain) state. (b) Squared overlaps of the QP states for the full Brillouin zone (BZ). (c) Squared overlaps of the QH states for the full BZ. (d) Ratio $W/\Delta$ for the $\nu=1/3$ QP band with $N_e=5$ (blue crosses), where $W=E_{\mathrm{QP}}^{\max}-E_{\mathrm{QP}}^{\min}$ and $\Delta=E_{\mathrm{bulk}}^{\min}-E_{\mathrm{QP}}^{\min}$. Here $E_{\mathrm{bulk}}^{\min}$ is the lowest energy outside the QP band. The global spectral separation above the band is $\Delta-W$. Red dots show the squared overlap between the highest-energy CF QP state and the corresponding exact eigenstate. }
    \label{fig1}
\end{figure}

We apply this construction to the fermionic Jain states at
$\nu=1/3$, $2/5$, and $3/7$, as well as to the bosonic Laughlin state at $\nu=1/2$. Figure~\ref{fig1}(a) compares the resulting momentum-resolved CF states with the corresponding exact-diagonalization eigenstates of bare Coulomb interaction for a finite size system. Near the uniform-field limit, $\gamma=0$, the overlaps are close to unity. Upon increasing $\gamma$, the overlaps remain high over a finite interval, demonstrating that the Jain states are robust against moderate magnetic-field inhomogeneity. Beyond a characteristic modulation strength, however, the overlaps decrease rapidly and eventually become negligible. The modulation strength at which this crossover occurs decreases with $n$ along the sequence
$\nu=n/(2pn+1)$. This trend is consistent with the decreasing excitation gap of higher-order Jain states. 

{\it Quasiparticle and quasihole dispersions.---} The CF construction introduced above extends naturally to charged excitations through modifying the Slater determinant part. For a QP, the Slater determinant contains $n$ completely filled $\Lambda$ levels and one additional CF occupying an orbital in the next empty $\Lambda$ level. For a QH, one orbital is removed from the highest occupied $\Lambda$ level. We label the selected orbital in the auxiliary effective-field problem by $k_{CF}=0,1\cdots N_\phi^*-1$. After replacing the Slater determinant and applying the many-body momentum projector, we obtain QP and QH trial states with definite momentum $\mathbf K=(K_x,K_y)$. Their explicit forms are given in the Supplemental Material \cite{SM}.

The auxiliary label $k_{\rm CF}$ should not be confused with the physical many-body momentum $\mathbf K$. For a fixed $\mathbf K$, we find numerically that all nonvanishing momentum projections obtained from different choices of $k_{\rm CF}$ give the same normalized CF state. Thus, $k_{\rm CF}$ does not label distinct physical states after momentum projection. Some choices of $k_{\rm CF}$, however, may give a vanishing state because the unprojected CF wave function has no component in the selected momentum sector. For example, when $N_x=N_\phi^*$, a nonzero projection requires $k_{\rm CF}=K_x$. 

Figure~\ref{fig1} (b-c) compares the momentum-resolved CF states with the corresponding exact-diagonalization eigenstates throughout the electronic Brillouin zone. Although the overlaps vary with $\mathbf K$, they remain close to unity over most of the Brillouin zone for both QPs and QHs, demonstrating that the generalized CF construction accurately captures the charged excitations of the AC system.

\begin{figure}[tb]
    \centering    
    \includegraphics[width=0.98\linewidth]{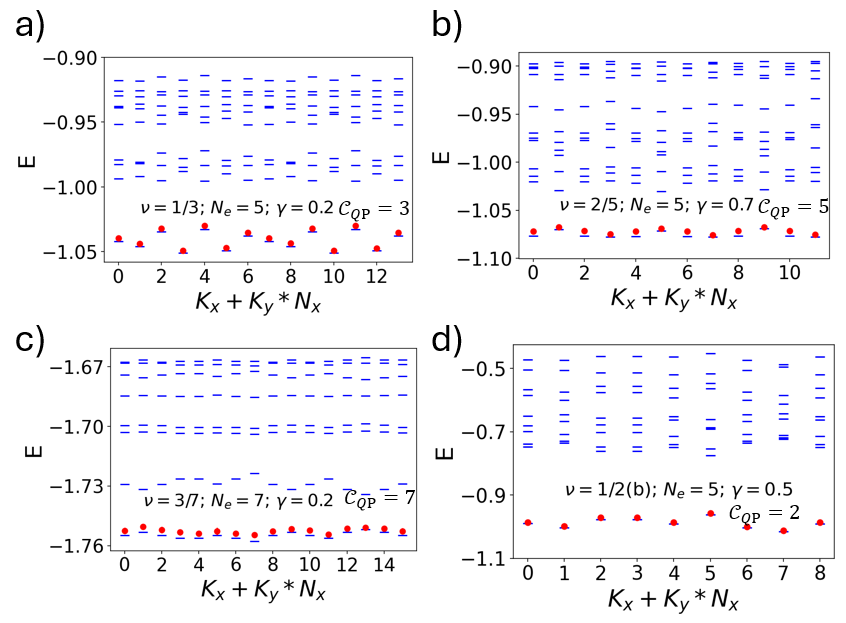}
    \caption{ (a-d) Quasiparticle energies for different fillings. The blue dashes are obtained from ED and red dots from CF. The filling factor, particle number, and values of $\gamma$ are labeled on the plots. We choose the lattice $(N_x,N_y)=$ $(7,2)$ (a), $(4,3)$ (b), $(4,4)$ (c) and $(3,3)$ (d). All energies are quoted in $e^2/\epsilon \ell_0$. We also label the QP band Chern number $\mathcal{C}_{\rm QP}$ in each panel, whose numerical results are shown in SM \cite{SM}.}
    \label{fig2}
\end{figure}

The energy of a single charged CF excitation directly probes the effective CF band structure, as illustrated schematically in Fig.~\ref{fig0}(d) and \ref{fig0}(e). In the uniform-field limit, $\gamma=0$, the AC band reduces to the lowest Landau level. Continuous magnetic-translation symmetry then relates different guiding-center positions and enforces a momentum-independent excitation energy. Accordingly, the effective CF $\Lambda$ levels are flat, which is not necessarily true if $|\gamma|>0$.

We calculate these dispersions using both exact diagonalization and the variational energies of the momentum-projected CF states. As shown in Fig.~\ref{fig2}, the two approaches agree quantitatively throughout the Brillouin zone, providing a stringent test of our CF construction. Additional variational CF dispersions for larger systems beyond those accessible to our ED calculations are presented in the Supplemental Material \cite{SM}.

The resulting QP and QH bands acquire finite bandwidths for any nonzero modulation, which must arise entirely from the interplay between electron-electron interactions and the spatially nonuniform magnetic field, given that all single-particle states have zero kinetic energy. Importantly, the dispersion is reproduced by a wave function containing a single CF quasiparticle or quasihole and therefore does not require residual interactions between multiple CF excitations. 

The dispersion of the CF band also provides a useful measure of the stability of the FQH/FCI phase. We define the bandwidth as $W=E_{\mathrm{QP}}^{\max}-E_{\mathrm{QP}}^{\min}$ and the energy separation as $\Delta=E_{\mathrm{bulk}}^{\min}-E_{\mathrm{QP}}^{\min}$, where $E_{\mathrm{bulk}}^{\min}$ is the lowest energy outside the QP band in the same charged sector. The global spectral separation above the QP band is therefore $\Delta-W$. Fig.~\ref{fig1}(d) shows that $W/\Delta$ increases approximately linearly with the modulation strength $\gamma$ at $\nu=1/3$; results for other fillings are given in the Supplemental Material \cite{SM}. As this ratio approaches unity, the QP band loses global spectral isolation, accompanied by a rapid decrease in the overlap between the highest-energy CF QP state and the corresponding exact eigenstate. These observations identify a limit of defining the isolated single-CF-band and can be used as a measure of the transition away from FQH/FCI phase as the inhomogeneity is increased, while more rigorous criteria need to be set by the ground state's topological response.

Dispersive Laughlin quasiholes in Aharonov--Casher bands were previously demonstrated in Refs.~\cite{yan2025anyon,iyer2026dispersion}. Our results generalize these developments by constructing and quantitatively validating both quasiparticle and quasihole bands \cite{SM} for the Jain sequence, thus revealing not only the anyon dispersion but also the dispersive CF band structure. The effective CF $\Lambda$ levels in an AC band should consequently be viewed not as flat Landau levels, but as emergent dispersive bands. 

\textit{Topology of the quasiparticle band.---}
The isolated QP band contains one normalized many-body state
$|\Psi_{\mathbf k}\rangle$ at each crystal momentum $\mathbf k$ in the
electronic Brillouin zone (BZ), with the physical torus boundary twists
held fixed. We characterize its topology using density-operator links
between neighboring momentum states. The phase accumulated around an
elementary momentum-space plaquette defines its Berry flux
$\Omega_{\mathrm{QP}}(\mathbf k)$, as illustrated in Fig.~\ref{fig0}(f). Summing these
fluxes over the electronic BZ gives the QP band Chern number Eq.~\ref{CQP}.
The link definition, its relation to overlaps of many-body periodic
parts, and the treatment of BZ boundaries are given in the End Matter.

For a single QP of the fermionic Jain state at
$\nu=n/(2pn+1)$, we obtain
\begin{equation}
 \mathcal{C}_{\mathrm{QP}}=2pn+1,\qquad
 \Phi_{\mathrm{BZ}}=2\pi (2pn+1).
 \label{eq:qp-main-result}
\end{equation}
In the uniform-field limit, this follows from the exact center-of-mass
magnetic-translation algebra and the single-QP counting relation
$(2pn+1)N_e-nN_\phi=1$, as shown in the End Matter.
The resulting integer is preserved upon continuously increasing
$\gamma$, provided the QP branch remains isolated, all density links
remain nonzero, and no plaquette phase crosses the principal-branch
boundary. These conditions and the finite-mesh resolution requirement
are stated explicitly in the End Matter.

Our CF calculations give $\mathcal{C}_{\mathrm{QP}}=3,5,7$ for
$\nu=1/3,2/5,3/7$, respectively, and $\mathcal{C}_{\mathrm{QP}}=2$ for the
bosonic $\nu=1/2$ state. We also confirm these results with the exact eigenstates obtained from ED. The $\nu=1/3$ QP band obtained by exact
diagonalization in twisted bilayer MoTe$_2$ also has
$\mathcal{C}_{\mathrm{QP}}=3$~\cite{SM}. Thus, the total QP Berry flux over the
electronic BZ is $2\pi D$, whereas a single-electron Landau level has
total flux $2\pi$ and Chern number one in the same sign convention.
The QP-band invariant therefore retains the denominator of the parent
Jain filling even when field modulation makes the QP band dispersive.

\textit{Discussion.---}
We have developed a systematic composite-fermion wave-function framework for Jain states in AC bands. The resulting trial states accurately describe the ground states, quasiparticles, and quasiholes over a substantial range of magnetic-field inhomogeneity. Applying this construction to a single charged excitation reveals two features of the emergent CF band structure that go beyond the conventional picture of flat effective Landau levels: the quasiparticle and quasihole bands acquire a finite dispersion, and the isolated quasiparticle band carries a total Berry flux $2\pi(2pn+1)$ for filling $\nu=n/(2pn+1)$. Thus, a mobile CF excitation retains information about both the lattice-scale inhomogeneity and its parent state. Moreover, as long as the QP band remains isolated, its quantized Chern number $\mathcal{C}_{\rm QP}=2pn+1$ provides a complementary numerical diagnostic of the FCI phase.

Our construction provides a microscopic tool for studying mobile anyons throughout the Jain sequence. This opens a route toward analyzing collective phases at finite quasiparticle or quasihole density. Even in the conventional FQHE, where an isolated CF excitation has a flat dispersion, several anyons can bind into molecular states \cite{Gattu2025Molecular,Wang2026anyon,Xu2025dynamics}. A finite anyon bandwidth introduces an additional kinetic-energy scale and may therefore qualitatively alter the formation, internal structure, and stability of such bound states. Dispersive CF bands are also relevant to proposals for anyon superconductivity \cite{Laughlin1988Super,Laughlin1988The,Nosov_2026,han2026anyon,Pichler_2026}, whose realization depends sensitively on the competition between the anyon dispersion, interactions, and statistical gauge fields. Because our wave functions can be evaluated for different magnetic textures and interactions, they provide a basis for quantitatively investigating these phases in large systems and in models motivated by realistic Chern-band materials.

Our results suggest a heuristic indicator for the stability of Jain FCIs: the quasiparticle band remains well defined while its bandwidth is small compared with the gap to higher-energy excitations. As was reported in the studies of the fractional quantum Hall effect, the transport gap decreases approximately as $1/(2pn+1)$ along the Jain sequence $\nu=n/(2pn+1)$\cite{Jain1997,Park1999,Morf2002,Du1993,Rosales2021}, while the quasiparticle bandwidth grows with field inhomogeneity \cite{SM}. Their competition possibly makes higher-order Jain states progressively more fragile as $n$ increases. Confirming this proposed trend requires further finite-size studies of the gap above the ground-state manifold and its topological response across different torus geometries, given that the loss of an isolated quasiparticle band alone does not establish a transition of the parent FCI phase.

An important remaining question is how the emergent CF dispersion can be measured experimentally. In the Landau-level setting, the peaks in the local tunneling spectrum have been shown to correspond directly to the kinetic energies of CF excitations \cite{Pu2024Fingerprints}. A dispersive CF band should broaden these discrete spectral features into momentum-integrated bands, although a quantitative prediction will require calculating the full tunneling spectrum in the AC system. Alternatively, the anyon bandwidth may be extracted from impurity-induced quasiparticle interference, as proposed in Ref.~\cite{Wang2026measuring}. Recent scanning-tunneling measurements near impurities have also revealed anisotropic spatial structures associated with localized fractional excitations \cite{Park2026local}. Establishing how these structures encode the dispersion and geometry of the underlying CF bands is an important direction for future theoretical and experimental work.

{\it Acknowledgements.} The authors thank Ajit Balram for his insightful and helpful comments on our draft, and also Jainendra Jain, Kai Sun, Jiabin Yu and Guangyue Ji for fruitful discussions and their advice. We also would like to thank Yafis Barlas and Yuval Oreg for helpful discussions and collaboration on related projects. This work was supported by a seed grant initiative from the Center for Quantum Leaps at Washington University in St. Louis. We thank the developers of the Julia package FuzzifiED \cite{Zhou2025}, which we used to perform the exact diagonalization calculations. 

\providecommand{\textonehalf}{\ensuremath{\frac{1}{2}}}

\setcounter{equation}{0}
\renewcommand{\theequation}{E\arabic{equation}}
\renewcommand{\theHequation}{EM.\arabic{equation}}
\section*{End Matter}
\section{AC orbitals and crystal momentum on the torus}
\label{sec:ac-model-end-matter}

We use a rectangular torus with $L_\mu=N_\mu a_\mu$,
$N_\phi=N_xN_y$, and $L_xL_y=2\pi N_\phi\ell_0^2$, where
$\ell_0=B_0^{-1/2}$. The AC calculations use $L_x=L_y$; the formulas below
allow general rectangular aspect ratios. With $z=x+iy$ and
$\tau=iL_y/L_x$, choose $\mathbf A=(-\partial_yQ,\partial_xQ)$, where
\begin{align}
 Q(x,y)=-\frac{B_0y^2}{2}+B_0\gamma\left[
 \frac{\sin(2\pi x/a_x)}{(2\pi/a_x)^2}
 +\frac{\sin(2\pi y/a_y)}{(2\pi/a_y)^2}\right].
 \label{eq:model-Q}
\end{align}
Thus $\nabla^2Q=-B(\mathbf r)$, and the zero-mode condition gives
$\widetilde\psi_j(\mathbf r)=e^{Q(x,y)}f_j(z)$ with holomorphic $f_j$.

The magnetic translations along the coordinate axes are
\begin{equation}
 t_x(d)=e^{d\partial_x},\qquad
 t_y(d)=e^{idx/\ell_0^2}e^{d\partial_y}.
 \label{eq:model-magnetic-translations}
\end{equation}
For nonzero modulation, these are symmetries for lattice translations
that preserve the magnetic texture. The orbitals obey
$t_\mu(L_\mu)\widetilde\psi_j=e^{i\phi_\mu}\widetilde\psi_j$.
A basis satisfying these boundary conditions is~\cite{Pu2020Hall}
\begin{equation}
 f_j(z)=\vartheta\!\begin{bmatrix}
 (j+\phi_x/2\pi)/N_\phi\\[-1pt]
 -\phi_y/2\pi
 \end{bmatrix}
 \!\left(\frac{N_\phi z}{L_x}\,\middle|\,N_\phi\tau\right),
 \label{eq:model-theta-orbitals}
\end{equation}
where $j=0,\ldots,N_\phi-1$ and $\vartheta$ is the Jacobi theta function
with the displayed characteristics. These orbitals need not be orthogonal.

Because each primitive cell contains one flux quantum, $t_x(a_x)$ and
$t_y(a_y)$ commute. Their normalized joint eigenstates are
\begin{equation}
 |\psi_{\boldsymbol s}\rangle
 =\mathcal N_{\boldsymbol s}
 \sum_{r=0}^{N_y-1}e^{-2\pi i r s_y/N_y}
 |\widetilde\psi_{s_x+rN_x}\rangle,
 \label{eq:model-bloch-orbitals}
\end{equation}
with integer labels $0\le s_\mu<N_\mu$ and normalization
$\mathcal N_{\boldsymbol s}$. They satisfy
\begin{equation}
 t_\mu(a_\mu)|\psi_{\boldsymbol s}\rangle
 =e^{i(2\pi s_\mu+\phi_\mu)/N_\mu}
 |\psi_{\boldsymbol s}\rangle.
 \label{eq:model-bloch-eigenvalues}
\end{equation}
Distinct eigenvalue pairs ensure orthogonality. The corresponding
single-particle wave vector is
$\kappa_\mu=(2\pi s_\mu+\phi_\mu)/L_\mu$, modulo $2\pi/a_\mu$.

For $N_e$ particles, $t_{\mathrm{CM}}(\mathbf a)=\prod_{i=1}^{N_e}t_i(\mathbf a)$.
The integer $S_\mu=\sum_{i=1}^{N_e}s_{\mu,i}\pmod{N_\mu}$ labels a many-body
momentum sector. The center-of-mass translation eigenvalue and the dimensional
momentum used in the QP-band construction are
\begin{align}
 t_{\mathrm{CM}}(a_\mu\hat{\mathbf e}_\mu)|\Psi_{\mathbf k}\rangle
 &=e^{i(2\pi S_\mu+N_e\phi_\mu)/N_\mu}|\Psi_{\mathbf k}\rangle,
 \nonumber\\
 k_\mu&=\frac{2\pi S_\mu+N_e\phi_\mu}{L_\mu}
 \pmod{\frac{2\pi}{a_\mu}}.
 \label{eq:model-many-body-momentum}
\end{align}
The physical twists $\phi_\mu$ are held fixed when traversing this
electronic Brillouin zone.

\section{Definition and quantization of the quasiparticle-band Chern number}
\label{sec:qp-end-matter}

\textit{Density links and plaquette flux.---}
Let $|\Psi_{\mathbf k}\rangle$ be the normalized state of an isolated
single-QP band at momentum $\mathbf k$. The physical boundary twists
are fixed throughout. On the $N_x\times N_y$ electronic momentum mesh,
write $\mathbf q_\mu=(2\pi/L_\mu)\hat{\mathbf e}_\mu$ and define
\begin{align}
 \rho(\mathbf q)&=\sum_{j=1}^{N_e}e^{i\mathbf q\cdot\mathbf r_j},
 \nonumber\\
 M_\mu(\mathbf k)&=
 \langle\Psi_{\mathbf k+\mathbf q_\mu}|
 \rho(\mathbf q_\mu)|\Psi_{\mathbf k}\rangle,
 \nonumber\\
 \mathcal U_\mu(\mathbf k)&=
 \frac{M_\mu(\mathbf k)}{|M_\mu(\mathbf k)|}.
 \label{eq:qp-links}
\end{align}
Related density-operator constructions have
been used to evaluate composite-fermion Berry
phases~\cite{Wang2019Lattice,Wang2019Dirac,Ji2020Berry,Geraedts2018Berry}.

For nonzero links, the counterclockwise plaquette product is
\begin{align}
 W_\square(\mathbf k)
 =\mathcal U_y(\mathbf k)^{-1}\mathcal U_x(\mathbf k+\mathbf q_y)^{-1}\mathcal U_y(\mathbf k+\mathbf q_x)\mathcal U_x(\mathbf k).
 \label{eq:qp-wilson}
\end{align}
We use the orientation convention
\begin{align}
 \Omega_{\mathrm{QP}}(\mathbf k)&=-\operatorname{Arg}W_\square(\mathbf k),
 \nonumber\\
 \mathcal{C}_{\mathrm{QP}}&=\frac{1}{2\pi}\sum_{\mathbf k}\Omega_{\mathrm{QP}}(\mathbf k),
 \label{eq:qp-flux}
\end{align}
where $\operatorname{Arg}$ is the principal argument. This convention
assigns positive $\mathcal{C}_{\mathrm{QP}}$ to the QP band for the field
$\mathbf B=-B_0\hat{\mathbf z}$ used here.
Under $|\Psi_{\mathbf k}\rangle\mapsto
e^{i\theta(\mathbf k)}|\Psi_{\mathbf k}\rangle$, a link acquires the
phase $e^{i[\theta(\mathbf k)-\theta(\mathbf k+\mathbf q_\mu)]}$,
which cancels around the plaquette. With consistent periodic
identifications, every link also cancels from
$\prod_{\mathbf k}W_\square(\mathbf k)$, making
Eq.~\eqref{eq:qp-flux} an integer-valued, gauge-invariant lattice
construction~\cite{Fukui2005Chern}.

\textit{Periodic-part representation.---}
The insertion of $\rho(\mathbf q)$ has an exact overlap interpretation.
Exchange symmetry makes the matrix elements of
$e^{i\mathbf q\cdot\mathbf r_j}$ identical for all $j$. Hence, on defining
$|u_{\mathbf k}\rangle=e^{-i\mathbf k\cdot\mathbf r_1}
|\Psi_{\mathbf k}\rangle$, one obtains
\begin{align}
 &\langle\Psi_{\mathbf k+\mathbf q}|
 \rho(\mathbf q)|\Psi_{\mathbf k}\rangle
 \nonumber\\
 &\qquad=N_e\langle\Psi_{\mathbf k+\mathbf q}|
 e^{i\mathbf q\cdot\mathbf r_1}|\Psi_{\mathbf k}\rangle
 \nonumber\\
 &\qquad=N_e\langle u_{\mathbf k+\mathbf q}|u_{\mathbf k}\rangle.
 \label{eq:qp-periodic-overlap}
\end{align}
Taking $\mathbf k$ to include the fixed twist offset, these states obey
$t_{\mathrm{CM}}(\mathbf a)|u_{\mathbf k}\rangle=|u_{\mathbf k}\rangle$
for primitive lattice vectors $\mathbf a$. Thus, the normalized density
links are precisely normalized overlaps of many-body periodic parts.
The change of representation acts
in the labelled-coordinate Hilbert space and transforms the exchange
constraint along with the wave function; choosing another particle
gives the same overlaps.

For a reciprocal lattice vector $\mathbf G$, the sewing relation
$|\Psi_{\mathbf k+\mathbf G}\rangle
=e^{i\chi_{\mathbf G}(\mathbf k)}|\Psi_{\mathbf k}\rangle$ becomes
\begin{equation}
 |u_{\mathbf k+\mathbf G}\rangle
 =e^{-i\mathbf G\cdot\mathbf r_1}
 e^{i\chi_{\mathbf G}(\mathbf k)}|u_{\mathbf k}\rangle.
 \label{eq:qp-sewing}
\end{equation}
Boundary links retain this sewing and use the small physical transfer
$\mathbf q_\mu$, rather than the difference of wrapped momentum
labels. When the periodic parts sample a smooth isolated band on a
sufficiently resolved momentum mesh, this is the usual overlap
construction for its momentum-space line bundle.
The finite-mesh integer and its continuation below can be established
directly without assuming an interpolation between mesh points.
The parameter space is QP momentum, not the space of physical boundary
twists used to define the Hall conductance.

\textit{Uniform-field value.---}
For one QP of the fermionic Jain state at $\nu=n/D$, with
$D=2pn+1$, CF counting on the torus gives
$N_e=n(N_\phi-2pN_e)+1$. Consequently,
\begin{equation}
 DN_e-nN_\phi=1,\qquad
 \gcd(N_e,N_\phi)=1.
 \label{eq:qp-counting}
\end{equation}
At $\gamma=0$, the center-of-mass translations
\begin{align}
 T_\mu&=t_{\mathrm{CM}}\!\left(\frac{L_\mu}{N_\phi}\hat{\mathbf e}_\mu\right),
 \quad \mu=x,y,
 \nonumber\\
 T_xT_y&=\omega T_yT_x,
 \quad \omega=e^{2\pi iN_e/N_\phi},
 \label{eq:qp-translations}
\end{align}
commute with the Hamiltonian~\cite{Haldane1985Many,Bernevig2012Emergent}.
The sign in Eq.~\eqref{eq:qp-translations} follows from the magnetic
translations used in the main text for $\mathbf B=-B_0\hat{\mathbf z}$.
Coprimality makes each irreducible multiplet $N_\phi$ dimensional.
Let $P$ project onto one isolated such multiplet representing the QP
band. For $N_\phi=N_xN_y$, the commuting translations
$T_x^{N_y}$ and $T_y^{N_x}$ give one state per electronic momentum
sector.

Define $A=P\bar\rho(\mathbf q_x)P$ and
$B=P\bar\rho(\mathbf q_y)P$, where $\bar\rho$ is the electronic
Landau-level-projected density. Translation covariance,
$t_{\mathrm{CM}}(\mathbf a)\bar\rho(\mathbf q)
t_{\mathrm{CM}}(\mathbf a)^{-1}
=e^{i\mathbf q\cdot\mathbf a}\bar\rho(\mathbf q)$, implies, with
$\zeta=e^{2\pi i/N_\phi}$,
\begin{align}
 T_xAT_x^{-1}&=\zeta A,& T_yAT_y^{-1}&=A,\nonumber\\
 T_xBT_x^{-1}&=B,& T_yBT_y^{-1}&=\zeta B.
 \label{eq:qp-covariance}
\end{align}
Equation~\eqref{eq:qp-counting} gives $\omega^D=\zeta$.
Therefore $AT_y^{-D}$ and $BT_x^D$ commute with both translations.
Schur's lemma yields, within the QP multiplet,
\begin{equation}
 A=c_xT_y^D,\qquad B=c_yT_x^{-D}.
 \label{eq:qp-schur}
\end{equation}
The coefficients $c_x,c_y$ are independent of the momentum label.
Assume $c_xc_y\ne0$, as required for the density links, and set
$Q_x=A/|c_x|$, $Q_y=B/|c_y|$. Their matrix elements implement the links
in Eq.~\eqref{eq:qp-links}. The counterclockwise product is
\begin{align}
 W_\square(\mathbf k)
 &=\langle\Psi_{\mathbf k}|
 Q_y^\dagger Q_x^\dagger Q_yQ_x|\Psi_{\mathbf k}\rangle
 \nonumber\\
 &=\langle\Psi_{\mathbf k}|
 T_x^DT_y^{-D}T_x^{-D}T_y^D|\Psi_{\mathbf k}\rangle
 \nonumber\\
 &=\omega^{-D^2}=e^{-2\pi iD/N_\phi}.
 \label{eq:qp-uniform-loop}
\end{align}
The phases of $c_x,c_y$ cancel. The same algebra applies to plaquettes
crossing the BZ boundary, with their sewing phases retained.

For $N_\phi>2D$, each principal plaquette flux is unambiguously
$2\pi D/N_\phi\in(0,\pi)$. Summing the $N_\phi$ plaquettes gives
\begin{equation}
 \Omega_{\mathrm{QP}}(\mathbf k)=\frac{2\pi D}{N_\phi},
 \qquad \Phi_{\mathrm{BZ}}=2\pi D,
 \qquad \mathcal{C}_{\mathrm{QP}}=D.
 \label{eq:qp-uniform-chern}
\end{equation}
The condition $N_\phi>2D$ resolves the flux on the principal branch;
on coarser meshes the loop phase alone specifies it only modulo
$2\pi$. For the bosonic $\nu=1/2$ state, single-vortex attachment gives
$N_e=(N_\phi-N_e)+1$, so the same proof applies with $D=2$.

\textit{Continuation to a modulated field.---}
Consider a continuous path from $\gamma=0$ at fixed boundary twists.
We follow one QP state in each momentum sector, separated by a
nonzero direct gap from the other states in that sector. Suppose
throughout this path that
\begin{equation}
 M_\mu(\mathbf k;\gamma)\ne0,\qquad
 W_\square(\mathbf k;\gamma)\ne-1
 \label{eq:qp-continuation-conditions}
\end{equation}
for every link and plaquette, with consistent BZ sewing. The first
condition makes the links continuous; the second prevents a
principal-argument branch crossing. Thus every
$\Omega_{\mathrm{QP}}(\mathbf k;\gamma)$ is continuous.
Since their sum divided by $2\pi$ is an integer, it cannot change:
\begin{equation}
 \mathcal{C}_{\mathrm{QP}}(\gamma)=\mathcal{C}_{\mathrm{QP}}(0)=D.
 \label{eq:qp-continuation}
\end{equation}
This proves the nonuniform-field result throughout the connected
parameter interval satisfying these conditions. In the AC
calculations, we follow the states adiabatically from $\gamma=0$ and
check for vanishing links and plaquette branch crossings.
For the finite-mesh construction, these checks accompany band
isolation; a spectral gap alone does not enforce them.

\clearpage
\onecolumngrid
\phantomsection
\pdfbookmark[0]{Supplemental Material}{arxiv-supplement}
\begin{center}
{\large\bfseries Supplemental Material for ``Composite fermions in ideal Chern bands''\par}
\medskip
Songyang Pu, Liangtao Peng, and Shaffique Adam\par
\smallskip
{\small Department of Physics, Washington University in St. Louis, St. Louis, Missouri 63130, USA\par}
\smallskip
{\small October 1, 2026\par}
\end{center}

\begin{center}
\begin{quote}
{\small In this supplementary material, we discuss the definition of the Aharonov-Casher band and its relations with the ideal Chern band, the many-body spectra from exact diagonalizations, the details of constructing composite fermion (CF) states and numerical results on the quasihole and quasiparticle bands, the symmetry in the many-body spectra and the definitions and numerical calculations of the Berry flux in both AC bands and twisted bilayer MoTe$_2$. 
}
\end{quote}
\end{center}

\setcounter{equation}{0}
\setcounter{figure}{0}
\setcounter{table}{0}
\setcounter{section}{0}

\renewcommand{\theequation}{S\arabic{equation}}
\renewcommand{\thefigure}{S\arabic{figure}}
\renewcommand{\thesection}{S\Roman{section}}

\renewcommand{\thetable}{S\arabic{table}}
\renewcommand{\theHequation}{SM.\arabic{equation}}
\renewcommand{\theHfigure}{SM.\arabic{figure}}
\renewcommand{\theHtable}{SM.\arabic{table}}
\renewcommand{\theHsection}{SM.\arabic{section}}
\renewcommand{\theHsubsection}{\theHsection.\arabic{subsection}}

\vspace{0cm}

\begin{center}
\textbf{CONTENTS}
\end{center}
\begingroup
\small
\setlength{\parindent}{0pt}
\setlength{\parskip}{0.25em}
\hyperref[arxiv-sm-section-1]{SI. Aharonov-Casher band}\hfill\pageref{arxiv-sm-section-1}\par
\hyperref[arxiv-sm-section-2]{SII. Exact diagonalization spectra}\hfill\pageref{arxiv-sm-section-2}\par
\hyperref[arxiv-sm-section-3]{SIII. CF construction and calculations}\hfill\pageref{arxiv-sm-section-3}\par
\hspace*{1.5em}\hyperref[arxiv-sm-section-3-1]{A. Construction of CF wave functions}\hfill\pageref{arxiv-sm-section-3-1}\par
\hspace*{1.5em}\hyperref[arxiv-sm-section-3-2]{B. More numerical results of QH and QP bands}\hfill\pageref{arxiv-sm-section-3-2}\par
\hyperref[arxiv-sm-section-4]{SIV. Symmetry of the spectra}\hfill\pageref{arxiv-sm-section-4}\par
\hspace*{1.5em}\hyperref[arxiv-sm-section-4-1]{A. Magnetic translation and combined inversion symmetry}\hfill\pageref{arxiv-sm-section-4-1}\par
\hspace*{1.5em}\hyperref[arxiv-sm-section-4-2]{B. Symmetry in the energy spectrum}\hfill\pageref{arxiv-sm-section-4-2}\par
\hyperref[arxiv-sm-section-5]{SV. Calculation of the Berry phase}\hfill\pageref{arxiv-sm-section-5}\par
\hyperref[arxiv-sm-section-6]{SVI. Continuum model and quasiparticle-band topology in twisted MoTe$_2$}\hfill\pageref{arxiv-sm-section-6}\par
\endgroup
\medskip
\section{Aharonov-Casher band}
\label{arxiv-sm-section-1}
We consider two-dimensional electrons moving in a periodically varying perpendicular magnetic field. The magnetic field is in the $-z$ direction ${\bf B}=\nabla\times{\bf A}=-{B}({\bf r})\hat{z}$. We use complex numbers for the coordinates and vector potential $z=x+iy$, $A=A_x+iA_y$, $\partial=2\partial/\partial \bar{z}=\partial_x+i\partial_y$. The kinetic momentum is $\Pi_{x,y}=-i\partial_{x,y}+A_{x,y}$, $\Pi=\Pi_x+i\Pi_y$.

In general, the band structure of noninteracting electrons is not flat, in contrast to the Landau levels in a homogeneous field. In the following two cases, an exactly flat zero-energy band exists, with orbitals satisfying $\Pi|\psi_0\rangle=0$. This is the Aharonov--Casher band\cite{Aharonov1979Ground}. The first case is that the electron is a Dirac fermion whose Hamiltonian is $H=v_F\begin{pmatrix}
    0&\bar{\Pi} \\
    \Pi&0
\end{pmatrix}$, and the zero mode is $(\psi_0,0)^T$. The second case is the non-relativistic electrons whose Hamiltonian is
\begin{align}
    H&={1\over 2m}\left[\Pi_x^2+\Pi_y^2-{gB(r)\over 2}\right]\nonumber \\
    &={1\over 2m}\left[\bar\Pi \Pi+i[\Pi_y,\Pi_x]-{gB(r)\over 2}\right]\nonumber \\
    &={1\over 2m}\left[\bar\Pi \Pi+B(r)(1-g/2)\right]
\end{align}
where we have used $i[\Pi_x,\Pi_y]=-B(r)$. If the gyromagnetic ratio $g=2$, the Hamiltonian becomes $H={1\over 2m}\bar\Pi \Pi$ and $|\psi_0\rangle$ are exactly the zero modes. The zero modes can be written as $\psi(z,\bar{z})=e^{Q(z,\bar{z})}f(z)$ where the magnetic potential is derived from $A=i\partial Q(z,\bar{z})$, since $\Pi\psi=e^Q(-i\partial Q-i\partial +i\partial Q)f(z)=0$. 

Refs. \cite{Wang2021Exact,Ledwith2023Vortex} have shown that the orbitals in a flat Chern band carries Chern number $|\mathcal{C}|=1$ which satisfies (i) the Berry curvature $\Omega({\bf k})$ is everywhere positive definite; and (ii) there exists a constant, positive, symmetric matrix $w^{ab}$ with unit determinant such that
$g^{ab}({\bf k})={1\over 2}w^{ab}\Omega({\bf k})$ or equivalently $\sum_{ab}w_{ab}g^{ab}({\bf k})=2\sqrt{\det g({\bf k})}=|\Omega({\bf k})|$ share the same form as the orbitals in the AC band, i.e.
\begin{equation}
\Psi_{\bf k}({\bf r})=N_{\bf k} e^{Q({\bf r})}f_{{\bf k}}(z),
\label{ideal}
\end{equation}
where both the normalization factor $N_{\bf k}$ and the exponential factor $Q({\bf r})$ encode the quantum geometry of the ideal Chern band. The normalization factor is directly determined by the Berry curvature through $w_{ab}\partial_k^a\partial_k^b\log N_{\bf k}=\Omega({\bf k})\mp 1$, where the sign $\mp$ corresponds to $\mathcal{C}=\pm1$. On the other hand, the exponential factor can be decomposed as $Q({\bf r})=Q_0+\delta Q({\bf r})$, where $Q_0$ is the gauge factor for a uniform field with one flux quantum per unit cell, i.e. $2\pi\ell_0^2=|{\bf a}_1\times {\bf a}_2|$ where $\ell_0$ is the magnetic length and $Q_0=-y^2/2\ell_0^2$ in Landau gauge, and $|e^{\delta Q({\bf r})}|^2=\sum_{\bf b}\tilde{w}_{\bf b}e^{i{\bf b}\cdot{\bf r}}$, given that $N_{\bf k}^{-2}=\sum_{\bf b}\tilde{w}_{\bf b}f_{\bf b}^{{\bf k}{\bf k}}$ where ${\bf b}$ is a reciprocal lattice vector. Here $f_{\bf b}^{{\bf k}_1{\bf k}_2}=\langle \Phi_{0\mathrm{LL},{\bf k}_1}|e^{i({\bf b}+{\bf k}_2-{\bf k}_1)\cdot{\bf r}}|\Phi_{0\mathrm{LL},{\bf k}_2}\rangle$ is the form factor for LLL. This is equivalent to the AC model. The uniform part of the magnetic field $B_0=\ell_0^{-2}$ is fixed by the size of the unit cell, and the nonuniform part is $\delta B({\bf r})=-(\partial_x^2+\partial_y^2)\delta Q({\bf r})$.

\section{Exact diagonalization spectra}
\label{arxiv-sm-section-2}
\begin{figure}[tb]
    \centering    
    \includegraphics[width=0.9\linewidth]{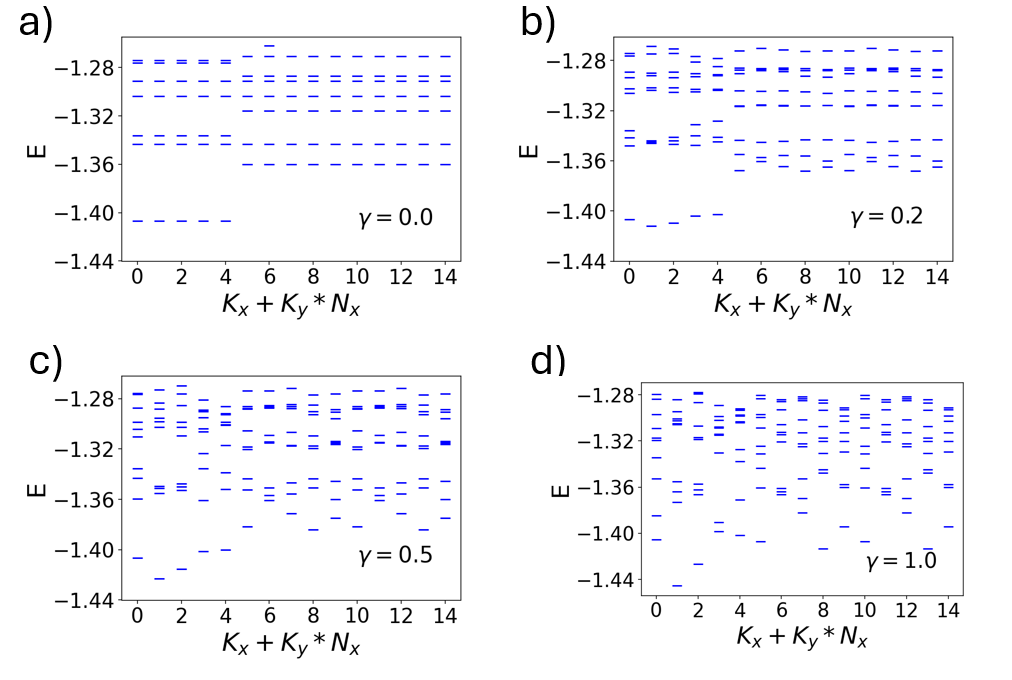}
    \caption{(a-d) Energy spectra for $N_e=6$, $\nu=2/5$ ground state sector obtained from exact diagonalizations with different $\gamma$. The gapped five-fold degenerate ground states can be clearly seen for $\gamma=0$ (a). While tuning on $\gamma$, the degeneracy is gradually broken, and the ground states are merging into the bulk of the spectrum, as shown in (b-d). }
    \label{EDGS}
\end{figure}

We first calculate the exact diagonalization (ED) spectra for the bare Coulomb interaction to show that the ground-state degeneracy is gradually lifted when increasing $\gamma$, as shown in Fig.~\ref{EDGS}. We show the spectrum for $N_e=6$ electrons at filling $\nu=2/5$ with different $\gamma$. At $\gamma=0$, the AC band reduces to the lowest Landau level, and the spectrum exhibits the expected fivefold center-of-mass degeneracy. Increasing $\gamma$ breaks the continuous magnetic-translation symmetry down to the lattice subgroup and splits the five low-energy states. For sufficiently strong modulation, this low-energy manifold merges with the higher excited states, signaling the loss of the clearly identifiable $\nu=2/5$ FCI phase.

\begin{figure}[tb]
    \centering    
    \includegraphics[width=0.9\linewidth]{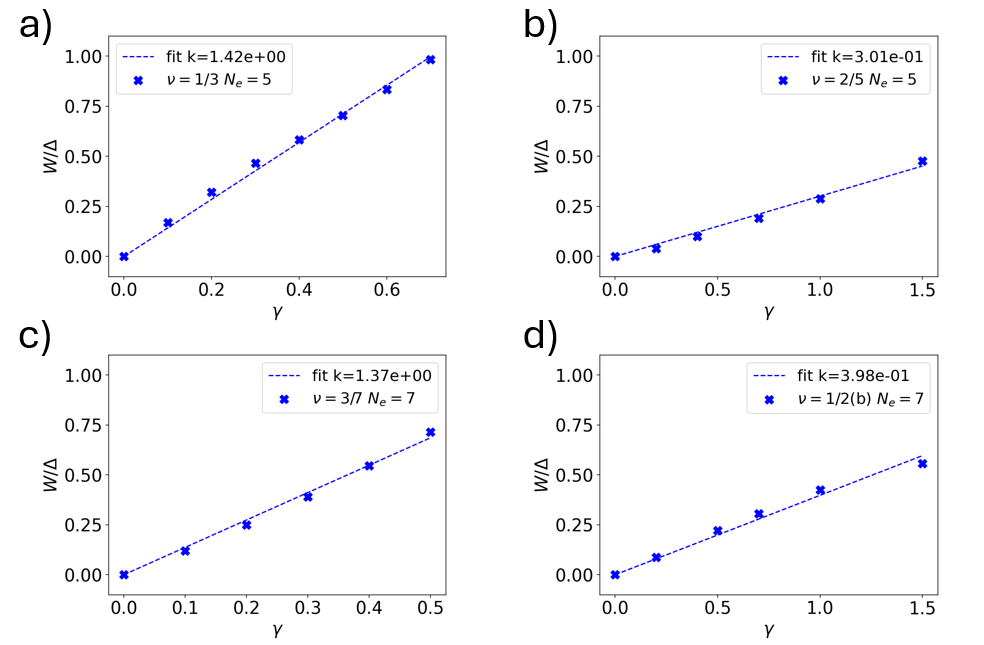}
    \caption{(a-d) The ratios of the bandwidth ($W$) and energy separation ($\Delta$) as a function of $\gamma$. The bandwidth is defined as the difference between the highest and lowest energy of the QP band. The energy separation is defined as the difference between the lowest energy above the QP band and the lowest energy of the QP band. We find the ratio increases linearly with $\gamma$.}
    \label{BW}
\end{figure}

Similarly, the QP band is exactly flat if $\gamma=0$. When $\gamma$ increases, the bandwidth increases, and finally the QP band merges with the bulk of the spectrum. We use ED to calculate the bandwidth, which is the difference between the highest and lowest energy of the QP band, and the energy separation, which is the difference between the lowest energy above the QP band and the lowest energy of the QP band, as a function of $\gamma$, as shown in Fig.~\ref{BW}, and find that their ratio increases linearly with $\gamma$ for all fillings we calculated (i.e. $\nu=1/3,2/5,3/7$ fermionic states and $\nu=1/2$ bosonic state). As $W/\Delta$ approaches unity, the global spectral separation $\Delta-W$ closes and the QP band ceases to be globally isolated from the higher-energy spectrum.

\section{CF construction and calculations}
\label{arxiv-sm-section-3}
\subsection{Construction of CF wave functions}
\label{arxiv-sm-section-3-1}
In this section, we discuss in detail the CF construction for AC band in the torus geometry, which largely follows the CF construction in the LLL developed in Ref.~\cite{Pu17}. As discussed in the main text, the generalization to AC is straightforward if we ignore the lattice magnetic translation symmetry. We just need to put the right magnetic potential $\sum_iQ_i$ in the Gaussian factor. The translation symmetry is then restored by a momentum projection operator as discussed in the main text. 

We model the system on a torus with boundaries $L_x$ and $L_y$ and periodic boundary conditions $t(L_\alpha)|\Psi\rangle=e^{i\phi_\alpha}|\Psi\rangle$, $\alpha=x,y$. Suppose the total magnetic flux number is $N_\phi$, the single-electron orbital in the lowest Landau level can be written as 
\begin{equation}
    \psi_k^{(0)}(x,y)=e^{Q(x,y)}f_k^{(0)}(z)
\end{equation}
\begin{equation}
f_k^{(0)}(z)=\elliptic[]{{k\over N_\phi}+{\phi_x\over 2\pi N_\phi}}{-{\phi_y\over 2\pi}}{N_\phi z/L_x}{N_\phi\tau}
\end{equation}
where $z=x+iy$, $Q(x,y)=-B_0{y^2\over 2}$ and $\elliptic[]{a}{b}{z}{\tau}\equiv \sum_{n=-\infty}^\infty e^{i\pi(n+a)^2\tau}e^{i2\pi(n+a)(z+b)}$ is the Jacobi-Theta function with rational characteristics \cite{Mumford07}. When applied to the AC model, $Q$ needs to be replaced by the magnetic potential defined by the nonuniform field, in our case $Q(x,y)=-B_0\left({y^2\over 2}-{\gamma\over \left(2\pi\over a_x\right)^2}\sin{2\pi x\over a_x}-{\gamma\over \left(2\pi\over a_y\right)^2}\sin{2\pi y\over a_y}\right)$. These orbitals satisfy 
\begin{align}
    t(L_x/N_\phi)|\psi_k^{(0)}\rangle&=e^{i{\phi_x+2\pi k\over N_\phi}}|\psi_k^{(0)}\rangle\\
    t(L_y/N_\phi)|\psi_k^{(0)}\rangle&=e^{i{\phi_y\over N_\phi}}|\psi_{k+1}^{(0)}\rangle
\end{align}
in a uniform field $\gamma=0$. The orbitals in the $n$th Landau level can be obtained from the ladder operator $\psi_k^{(n)}=e^{Q}f_k^{(n)}=e^{Q}\left(a_f^\dagger\right)^n f_k^{(0)}$ where $a_f^\dagger\equiv \sqrt{2}\ell_0\left({\bar{z}-z\over 2\ell_0^2}-\partial_z\right)$.

The CF wave function can be written as $\Psi_{CF}=\mathcal{P}_{\rm AC}\Psi_n\Psi_1^{2p}$, where $\mathcal{P}_{\rm AC}$ is the AC band projection operator that enforces $\Pi_i\mathcal{P}_{\rm AC}|\Psi\rangle=0$, $\Psi_n$ is the Slater determinant $\Psi_n={\rm Det}\left[\psi_k^{(m)}(z_i)\right]$ with $n$ effective LLs filled (or the corresponding CF qp or qh wave functions as discussed below), and its special case for $n=1$ $\Psi_1$ can be written as
\begin{equation}
    \Psi_1=e^{\sum_i Q_i}\vartheta \begin{bmatrix}
(N_e-1)/2 \\ (N_e-1)/2
\end{bmatrix}\Bigg({\frac{Z}{L_x}} \Bigg | \tau \Bigg) \prod_{i<j}  \vartheta 
\begin{bmatrix}
\frac{1}{2} \\ \frac{1}{2}
\end{bmatrix}
\Bigg( \frac{z_i-z_j}{L_x}\Bigg | \tau \Bigg )
\end{equation}
where $Z=\sum_i z_i$. The exact LLL projection makes the evaluation of CF wave functions very expensive. Instead, we implement the Jain-Kamilla projection \cite{Jain97,Jain1997}. The resulting ground state wave functions for $\nu=n/(2pn+1)$ is
\begin{equation}
    \Psi_{\rm CF}^{\nu=n/(2pn+1)}=e^{\sum_i Q_i}\left(\vartheta \begin{bmatrix}
(N_e-1)/2 \\ (N_e-1)/2
\end{bmatrix}\Bigg({\frac{Z}{L_x}} \Bigg | \tau \Bigg) \right)^{2p}\chi_{\rm GS}^{\nu=n/(2pn+1)}
\end{equation}
\begin{equation}
    \chi_{\rm GS}^{\nu=n/(2pn+1)}=
        \begin{vmatrix}
        \hat{g}^{(0)}_0(z_1)J_1^p\cdots \hat{g}^{(0)}_0(z_{N_e})J_{N_e}^p\\
        \vdots\\
        \hat{g}^{(0)}_{N_\phi^*-1}(z_1)J_1^p\cdots \hat{g}^{(0)}_{N_\phi^*-1}(z_{N_e})J_{N_e}^p\\
        \hat{g}^{(1)}_0(z_1)J_1^p\cdots \hat{g}^{(1)}_0(z_{N_e})J_{N_e}^p\\
        \vdots\\
        \hat{g}^{(n-1)}_{N_\phi^*-1}(z_1)J_1^p\cdots \hat{g}^{(n-1)}_{N_\phi^*-1}(z_{N_e})J_{N_e}^p\\
    \end{vmatrix} 
\end{equation}
\begin{equation}
J_i\equiv \prod_{j\neq i}  \vartheta 
\begin{bmatrix}
\frac{1}{2} \\ \frac{1}{2}
\end{bmatrix}
\Bigg( \frac{z_i-z_j}{L_x}\Bigg | \tau \Bigg ) 
\end{equation}
where $\hat{g}_k^{(m)}$ is an operator derived from $f_k^{(m)}$ through the JK projection, whose details can be found in Ref.~\cite{Pu17}. Below we list the explicit expressions for the first few $\Lambda$ levels, where $N_\phi^*=N_\phi-2pN_e$ and the normalization factor are omitted
\begin{align}
\hat{g}_k^{(0)}(z)&=f_k^{(0)}(z)\nonumber \\
\hat{g}_k^{(1)}(z)&=(N_\phi^*-N_\phi)\partial_zf_k^{(0)}(z)+2N_\phi^*f_k^{(0)}(z)\partial_z\nonumber \\
\hat{g}_k^{(2)}(z)&=(N_\phi^*-N_\phi)^2\partial_z^2f_k^{(0)}(z)+4N_\phi^*(N_\phi^*-N_\phi)\partial_zf_k^{(0)}(z)\partial_z+4N_\phi^{*2}f_k^{(0)}(z)\partial_z^2\nonumber \\
\hat{g}_k^{(3)}(z)&=(N_\phi^*-N_\phi)^3\partial_z^3f_k^{(0)}(z)+6N_\phi^*(N_\phi^*-N_\phi)^2\partial_z^2f_k^{(0)}(z)\partial_z+12N_\phi^{*2}(N_\phi^*-N_\phi)\partial_zf_k^{(0)}(z)\partial_z^2+8N_\phi^{*3}f_k^{(0)}(z)\partial_z^3\nonumber \\
.\end{align}
Similarly, the quasiparticle wave function can be obtained by making the replacement $ \chi_{\rm GS}^{\nu=n/(2pn+1)}\rightarrow  \chi_{\rm QP}^{\nu=n/(2pn+1)}$ 
\begin{equation}
    \chi_{\rm QP}^{\nu=n/(2pn+1)}=
        \begin{vmatrix}
        \hat{g}^{(0)}_0(z_1)J_1^p\cdots \hat{g}^{(0)}_0(z_{N_e})J_{N_e}^p\\
        \vdots\\
        \hat{g}^{(0)}_{N_\phi^*-1}(z_1)J_1^p\cdots \hat{g}^{(0)}_{N_\phi^*-1}(z_{N_e})J_{N_e}^p\\
        \hat{g}^{(1)}_0(z_1)J_1^p\cdots \hat{g}^{(1)}_0(z_{N_e})J_{N_e}^p\\
        \vdots\\
        \hat{g}^{(n-1)}_{N_\phi^*-1}(z_1)J_1^p\cdots \hat{g}^{(n-1)}_{N_\phi^*-1}(z_{N_e})J_{N_e}^p\\
        \hat{g}^{(n)}_{k_{CF}}(z_1)J_1^p\cdots \hat{g}^{(n)}_{k_{CF}}(z_{N_e})J_{N_e}^p\\
    \end{vmatrix} 
\end{equation}
which creates a qp on the $n$th effective LL with CF momentum $k_{CF}$ and the quasihole wave function is
\begin{equation}
    \chi_{\rm QH}^{\nu=n/(2pn+1)}=
        \begin{vmatrix}
        \hat{g}^{(0)}_0(z_1)J_1^p\cdots \hat{g}^{(0)}_0(z_{N_e})J_{N_e}^p\\
        \vdots\\
        \hat{g}^{(0)}_{N_\phi^*-1}(z_1)J_1^p\cdots \hat{g}^{(0)}_{N_\phi^*-1}(z_{N_e})J_{N_e}^p\\
        \hat{g}^{(1)}_0(z_1)J_1^p\cdots \hat{g}^{(1)}_0(z_{N_e})J_{N_e}^p\\
        \vdots\\
        \hat{g}^{(n-1)}_{k_{CF}-1}(z_1)J_1^p\cdots \hat{g}^{(n-1)}_{k_{CF}-1}(z_{N_e})J_{N_e}^p\\
        \hat{g}^{(n-1)}_{k_{CF}+1}(z_1)J_1^p\cdots \hat{g}^{(n-1)}_{k_{CF}+1}(z_{N_e})J_{N_e}^p\\
        \vdots\\
        \hat{g}^{(n-1)}_{N_\phi^*-1}(z_1)J_1^p\cdots \hat{g}^{(n-1)}_{N_\phi^*-1}(z_{N_e})J_{N_e}^p\\
    \end{vmatrix} 
\end{equation}
which creates a QH on the $n-1$th effective LL with CF momentum $k_{CF}$.

The constructed CF states above satisfy the periodic boundary conditions $t_i(L_{x(y)})|\Psi_{\rm CF}\rangle=e^{i\phi_{1(2)}}|\Psi_{\rm CF}\rangle$. As mentioned in the main text and the symmetry section below, the many-body states are expected to be the eigenstates of $t_{CM}(a_{x(y)})$. While the original CF states do not satisfy this, we can define a momentum projection operator
\begin{equation}
    \mathcal{P}_{(k_x,k_y)}=\sum_{m_1=0}^{N_x-1}\sum_{m_2=0}^{N_y-1}e^{-i2\pi\left({k_xm_1\over N_x}+{m_1\phi_1N_e\over2\pi N_x}+{k_ym_2\over N_y}+{m_2\phi_2N_e\over2\pi N_y}\right)}t_{\rm CM}^{m_1}(a_x)t_{\rm CM}^{m_2}(a_y)
\end{equation}
such that 
\begin{align}
    t_{CM}(a_{x})\mathcal{P}_{(k_x,k_y)}|\Psi_{\rm CF}\rangle&=e^{i{\phi_1N_e+2\pi k_x\over N_x}}\mathcal{P}_{(k_x,k_y)}|\Psi_{\rm CF}\rangle \nonumber \\
    t_{CM}(a_{y})\mathcal{P}_{(k_x,k_y)}|\Psi_{\rm CF}\rangle&=e^{i{\phi_2N_e+2\pi k_y\over N_y}}\mathcal{P}_{(k_x,k_y)}|\Psi_{\rm CF}\rangle \nonumber \\
.\end{align}

\subsection{More numerical results of QH and QP bands}
\label{arxiv-sm-section-3-2}
\begin{figure}[tb]
    \centering    
    \includegraphics[width=0.7\linewidth]{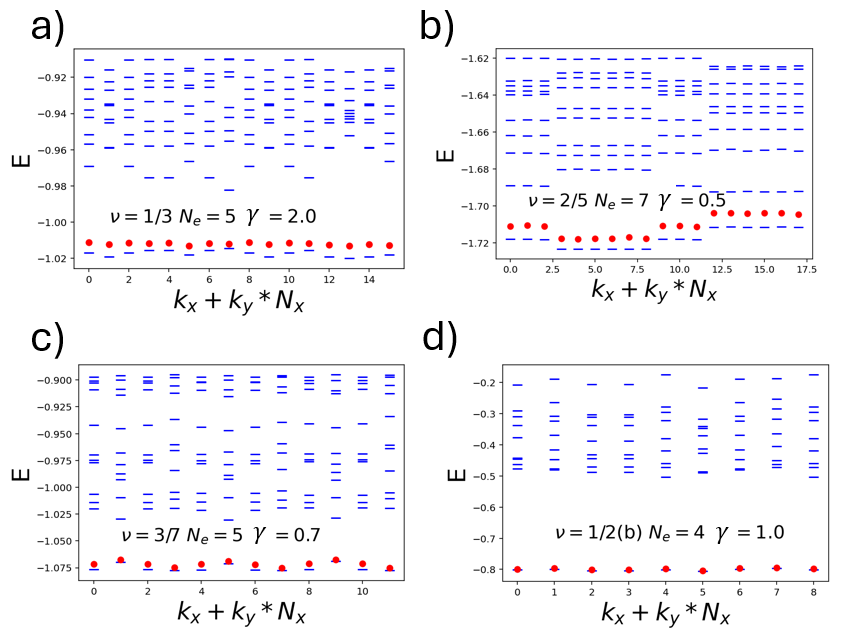}
    \caption{(a-d) Quasihole energies for different fillings. The blue dashes are obtained from ED and red dots from CF.}
    \label{qhband}
\end{figure}

We show the overlaps of the CF QH states and exact eigenstates in the main text. We also calculated the dispersion of the QH band and compared with ED results, as shown in Fig.~\ref{qhband}, which reveals a decent quantitative match.

\begin{figure}[tb]
    \centering    
    \includegraphics[width=0.7\linewidth]{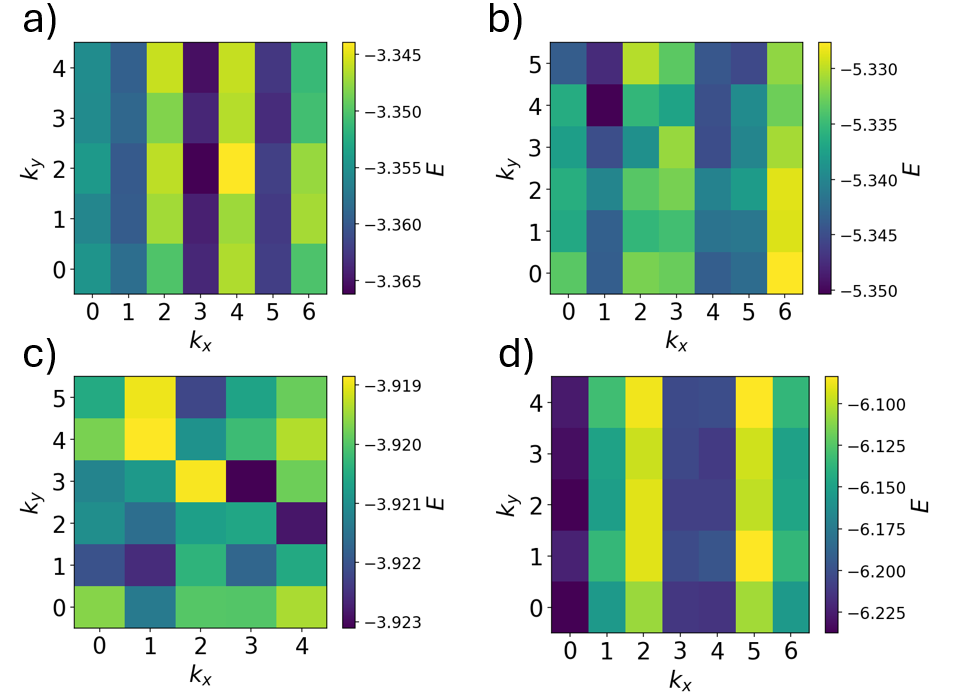}
    \caption{Quasiparticle band from CF calculations. (a) $\nu=1/3,N_e=12,N_x=7,N_y=5,\gamma=1.0$. (b) $\nu=2/5,N_e=17,N_x=7,N_y=6,\gamma=2.0$. (c)$\nu=3/7,N_e=13,N_x=5,N_y=6,\gamma=0.5$. (d) Bosonic $\nu=1/2,N_e=18,N_x=7,N_y=5,\gamma=1.0$.}
    \label{qpband}
\end{figure}

We also implement our CF construction on bigger systems to obtain the QP band in the full BZ, as shown in Fig.~\ref{qpband}.

\section{Symmetry of the spectra}
\label{arxiv-sm-section-4}
\subsection{Magnetic translation and combined inversion symmetry}
\label{arxiv-sm-section-4-1}
As mentioned above, our model has lattice magnetic translation symmetry spanned by $t_x\equiv t({\bf a}_x)$ and $t_y\equiv t({\bf a}_y)$. Besides the lattice magnetic translation symmetry, there is another symmetry in our model. While our model breaks the inversion symmetry $\mathcal{I}:{\bf r}\rightarrow-{\bf r}$ given that ${\bf B}({\bf r})\neq {\bf B}(-{\bf r})$, it still preserves a combined symmetry of inversion, half-lattice translation $T({a_x\over 2},{a_y\over 2}):(x,y)\rightarrow(x+{a_x\over 2},y+{a_y\over 2})$, and a gauge transformation, which is defined as 
\begin{align}
\mathcal{S}&=e^{i\chi_S({\bf r}_i)}T_i({a_x\over 2},{a_y\over 2})\mathcal{I};\nonumber \\
\chi_S({\bf r})&=-\frac{a_y}{2\ell_0^2}x;\nonumber \\
.\end{align}
To show $\mathcal{S}$ is a valid symmetry, we only need to confirm the one-body term $H={1\over 2m}\bar{\Pi}\Pi$ commutes with $\mathcal{S}$, since the two-body interaction obviously commutes with this spatial symmetry. We first look at $\Pi=\Pi_x+i\Pi_y=(-i\partial_x+A_x)+i(-i\partial_y+A_y)$. The transformation would act (${\bf d}\equiv({a_x\over 2},{a_y\over 2})$)
\begin{align}
    &e^{i\chi_S({\bf r})}T({\bf d})\mathcal{I}\left((-i\partial_x+A_x({\bf r}))+i(-i\partial_y+A_y({\bf r}))\right)\mathcal{I}^{-1}T(-{\bf d})e^{-i\chi_S({\bf r})}\nonumber \\
    =&e^{i\chi_S({\bf r})}T({\bf d})\left((i\partial_x+A_x(-{\bf r}))+i(i\partial_y+A_y(-{\bf r}))\right)T(-{\bf d})e^{-i\chi_S({\bf r})}\nonumber \\
    =&e^{i\chi_S({\bf r})}\left((i\partial_x+A_x(-{\bf r}+{\bf d}))+i(i\partial_y+A_y(-{\bf r}+{\bf d}))\right)e^{-i\chi_S({\bf r})}\nonumber \\
    =&-e^{i\chi_S({\bf r})}\left((-i\partial_x+A_x({\bf r})+\partial_x\chi_S)+i(-i\partial_y+A_y({\bf r})+\partial_y\chi_S)\right)e^{-i\chi_S(\mathbf r)}\nonumber \\
    =&-(-i\partial_x+A_x({\bf r}))-i(-i\partial_y+A_y({\bf r}))
\end{align}
therefore $\mathcal{S}\Pi\mathcal{S}^{-1}=-\Pi$. (In the fourth row, we used $-{\bf A}({\bf d}-{\bf r})={\bf A}({\bf r})+\nabla\chi_S$.) Similarly, one can prove $\mathcal{S}\bar{\Pi}\mathcal{S}^{-1}=-\bar{\Pi}$. Hence, $[H,\mathcal{S}]=0$.

We note that $\mathcal{S}$ does not commute with $t_x$ and $t_y$. Instead, we have 
\begin{equation}
\mathcal{S} t_x \mathcal{S}^{-1}=e^{-i a_xa_y/(2\ell_0^2)}t_x^{-1}.
\label{S18}
\end{equation}
and 
\begin{equation}
\mathcal{S} t_y \mathcal{S}^{-1}=e^{i a_xa_y/(2\ell_0^2)}t_y^{-1}.
\label{S19}
\end{equation}
 For one magnetic flux quantum per unit cell $a_xa_y=2\pi\ell_0^2$, both prefactors above equal $-1$. Therefore, we have $\mathcal{S} t_{x(y)} \mathcal{S}^{-1}=-t_{x(y)}^{-1}$.

\subsection{Symmetry in the energy spectrum}
\label{arxiv-sm-section-4-2}
The action of the symmetry on the spectrum depends on both the torus
boundary conditions and the momentum sector. Let
$\mathcal S=\prod_{i=1}^{N_e}S_i$ be the many-body transformation, and
let $t_{i,\mu}$ translate particle $i$ by one primitive lattice vector
in direction $\mu=x,y$. For one flux quantum per unit cell,
Eqs.~\ref{S18} and Eq.~\ref{S19} imply
\begin{equation}
 t_{i,\mu}\mathcal S=-\mathcal S\,t_{i,\mu}^{-1}.
 \label{eq:sym-primitive}
\end{equation}
We first determine the boundary conditions of the transformed state,
which depend on the parity of $N_\mu$, and then its momentum label,
which also depends on $N_e$.

\emph{Boundary twists.---}
Taking any one particle around the torus gives
\begin{equation}
 t_{i,\mu}^{N_\mu}|\Psi\rangle
 =e^{i\phi_\mu}|\Psi\rangle.
 \label{eq:sym-bc}
\end{equation}
Applying Eq.~\eqref{eq:sym-primitive} repeatedly yields
\begin{equation}
 t_{i,\mu}^{N_\mu}\mathcal S|\Psi\rangle
 =(-1)^{N_\mu}e^{-i\phi_\mu}\mathcal S|\Psi\rangle,
 \qquad
 \phi_\mu'=N_\mu\pi-\phi_\mu\pmod{2\pi}.
 \label{eq:sym-twist-map}
\end{equation}
Thus, for the initially periodic boundary conditions
$\phi_x=\phi_y=0$ considered below, the boundary condition in direction
$\mu$ remains periodic if $N_\mu$ is even and becomes antiperiodic
if $N_\mu$ is odd. This conclusion is independent of $N_e$ because
the boundary condition is imposed on each particle separately.
The transformation preserves the energy while mapping the Hamiltonian
with twists $\boldsymbol\phi$ to that with twists
$\boldsymbol\phi'$. It is a symmetry within a fixed boundary-twist
sector only when $\boldsymbol\phi'=\boldsymbol\phi$ modulo $2\pi$.

\emph{Momentum sectors.---}
Define the center-of-mass translation
$\mathcal T_\mu=\prod_{i=1}^{N_e}t_{i,\mu}$ and the integer momentum
label $K_\mu$ by
\begin{equation}
 \mathcal T_\mu|\Psi_{\mathbf K}^{(\boldsymbol\phi)}\rangle
 =\lambda_\mu|\Psi_{\mathbf K}^{(\boldsymbol\phi)}\rangle,
 \qquad
 \lambda_\mu=
 \exp\!\left[\frac{i(2\pi K_\mu+N_e\phi_\mu)}{N_\mu}\right].
 \label{eq:sym-momentum-convention}
\end{equation}
Here $K_\mu$ is defined modulo $N_\mu$ and uses the same convention
as the integer label $S_\mu$ in the End Matter.
Equation~\eqref{eq:sym-primitive} gives
\begin{equation}
 \mathcal T_\mu\mathcal S
 =(-1)^{N_e}\mathcal S\mathcal T_\mu^{-1},
 \qquad
 \lambda_\mu'=(-1)^{N_e}\lambda_\mu^{-1}.
 \label{eq:sym-cm-map}
\end{equation}
For even $N_e$, the translation eigenvalue is simply inverted;
for odd $N_e$, it also acquires a minus sign. To obtain the new integer
label, however, the twist contribution in
Eq.~\eqref{eq:sym-momentum-convention} must be retained.
Starting from $\phi_\mu=0$, choose the representative
$\phi_\mu'=\epsilon_\mu\pi$, where
$\epsilon_\mu=N_\mu\bmod 2$. Equating the transformed eigenvalues gives
\begin{equation}
 K_\mu'=\frac{N_e(N_\mu-\epsilon_\mu)}{2}-K_\mu
 \pmod{N_\mu}.
 \label{eq:sym-integer-map}
\end{equation}

\begin{table}[ht]
\centering
\begin{tabular}{|c|c|c|c|}
\hline
$N_\mu$ & $N_e$ & New twist $\phi_\mu'$ &
New integer label $K_\mu'$ (mod $N_\mu$) \\
\hline
Even & Even & $0$   & $-K_\mu$ \\
\hline
Even & Odd  & $0$   & $\dfrac{N_\mu}{2}-K_\mu$ \\
\hline
Odd  & Even & $\pi$ & $-\dfrac{N_e}{2}-K_\mu$ \\
\hline
Odd  & Odd  & $\pi$ & $\dfrac{N_\mu-N_e}{2}-K_\mu$ \\
\hline
\end{tabular}
\caption{The spectra symmetry for different $N_\mu$ and $N_e$.}
\label{table-sym}
\end{table}

The results are summarized in Table~\ref{table-sym}. For even $N_\mu$, the twist remains zero: even $N_e$ gives
$K_\mu'=-K_\mu$, whereas odd $N_e$ gives
$K_\mu'=N_\mu/2-K_\mu$.
For odd $N_\mu$, the new twist is $\pi$: even $N_e$ gives
$K_\mu'=-N_e/2-K_\mu$, whereas odd $N_e$ gives
$K_\mu'=(N_\mu-N_e)/2-K_\mu$.
In particular, the offset $-N_e/2$ for even $N_e$ originates from
the $N_e\phi_\mu'$ term in the translation eigenvalue; it is
consistent with the inversion of $\lambda_\mu$.

\begin{figure}[tb]
    \centering    
    \includegraphics[width=0.9\linewidth]{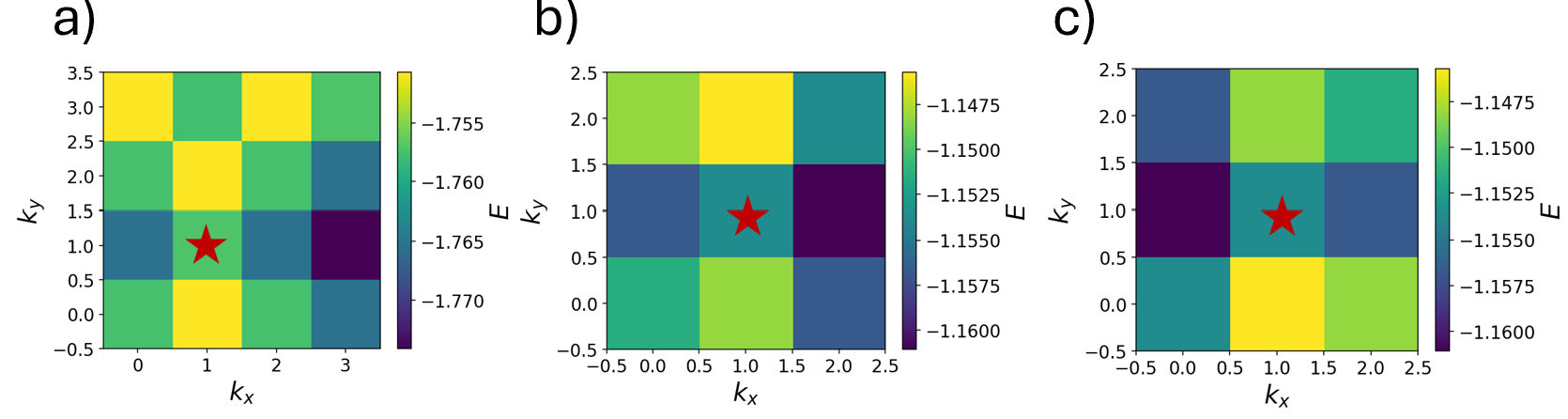}
    \caption{The exact ground state energies in each momentum sector. (a) When $N_e$ is odd while both $N_x$ and $N_y$ are even, the energies are the same for $(k_x,k_y)$ and $({N_{x}\over 2}-k_{x},{N_{y}\over 2}-k_{y})$. The red star marks the symmetry center. We chose $N_e=7$, $N_x=N_y=4$, $\gamma=1.0$. (b-c) When $N_e$, $N_x$ and $N_y$ are all odd, the states in $(k_x,k_y)$ with periodic boundary conditions $\phi_1=\phi_2=0$ (b) are degenerate with states in $((N_x-N_e)/2-K_x,(N_y-N_e)/2-K_y)$ with anti-periodic boundary conditions $\phi_1=\phi_2=\pi$ (c). We chose $N_e=5$, $N_x=N_y=3$, $\gamma=1.0$.} 
    \label{sym}
\end{figure}

We test these results by numerical calculations, as shown in Fig.~\ref{sym}. When both $N_x$ and $N_y$ are even, these relations connect sectors
within the same periodic-boundary spectrum. For example,
$N_e=7$ and $N_x=N_y=4$ give
$(K_x,K_y)\mapsto(2-K_x,2-K_y)$, as in Fig.~\ref{sym}(a).
If either $N_x$ or $N_y$ is odd, the relations instead connect
spectra with different boundary twists. For $N_e=5$ and
$N_x=N_y=3$, they give
$(K_x,K_y)\mapsto(2-K_x,2-K_y)$ together with
$(\phi_x,\phi_y):(0,0)\mapsto(\pi,\pi)$, as in
Figs.~\ref{sym}(b-c).
This change of boundary twists also occurs for even $N_e$;
in that case, inversion of the center-of-mass translation eigenvalue
alone does not imply a degeneracy within the same periodic-boundary
spectrum.

\section{Calculation of the Berry phase}
\label{arxiv-sm-section-5}

We first calculate the Berry flux for each plaquette and then add them. The Berry flux of a plaquette is (here ${\bf k}$ is the center-of-mass momenta of QP states)
\begin{equation}
    \Omega({\bf k})=-{\rm Arg}[M(\bf{k},\bf{k}+\delta \bf{k}_x)M(\bf{k}+\delta \bf{k}_x,\bf{k}+\delta \bf{k}_x+\delta \bf{k}_y)M^{-1}(\bf{k}+\delta \bf{k}_y,\bf{k}+\delta \bf{k}_x+\delta \bf{k}_y)M_y^{-1}(\bf{k},\bf{k}+\delta \bf{k}_y)]
    \label{EBC}
\end{equation}
where $M(\bf{k}_1,\bf{k}_2)\equiv \langle \Psi_{\bf{k}_2}|\sum_{i=1}^{N_e}e^{i(\bf{k}_2-\bf{k}_1)\cdot \bf{r}_i}|\Psi_{\bf{k}_1}\rangle$, $\delta \bf{k}_x=(2\pi/L_x,0)$ and $\delta \bf{k}_y=(0,2\pi/L_y)$. The operator $\sum_{i=1}^{N_e}e^{i(\bf{k}_2-\bf{k}_1)\cdot \bf{r}_i}$ is inserted so that the overlap will not be zero. It is similar to taking the periodic part of the Bloch states. A similar definition has earlier been used in Refs.~\cite{Wang2019Lattice,Wang2019Dirac,Ji2020Berry} to study the Berry phase of the composite fermion Fermi sea, where the QP excitation is gapless. In our calculation, each $|\Psi_{\bf K}\rangle$ is a QP state constructed by the CF theory. We use Metropolis Monte Carlo to calculate each $M(\bf{k}_1,\bf{k}_2)$. Finally, we do the summation $\sum_{\bf k}\Omega({\bf k})$ to obtain the total Berry phase.

\begin{figure}[tb]
    \centering    
    \includegraphics[width=0.98\linewidth]{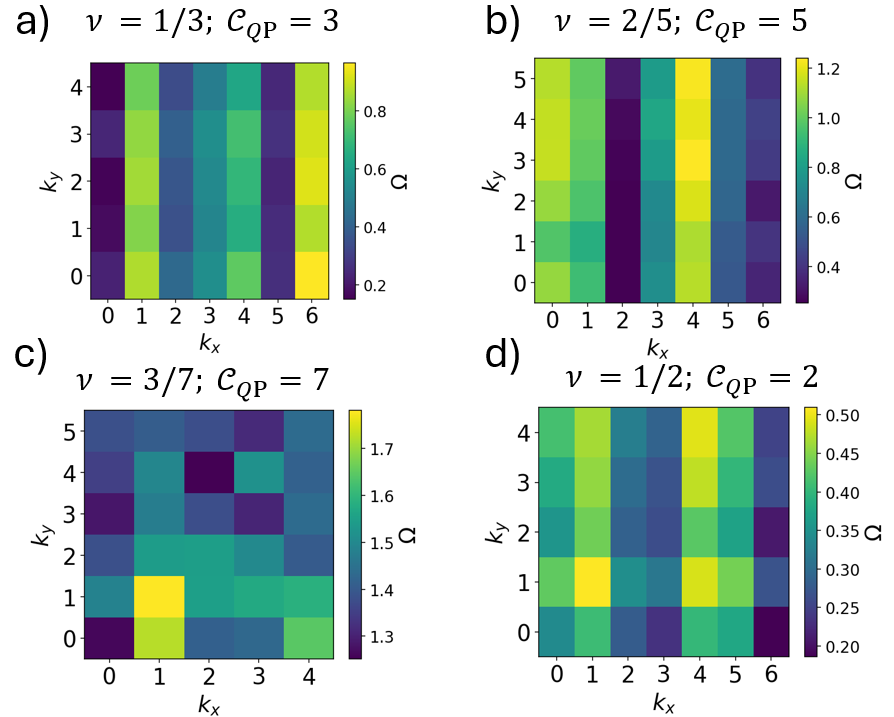}
    \caption{Berry flux of the QP band in each plaquette. (a) We choose $\nu=1/3,N_e=12,N_x=7,N_y=5,\gamma=1.0$ and find $\mathcal{C}_{QP}=3$. (b) We choose $\nu=2/5,N_e=17,N_x=7,N_y=6,\gamma=2.0$ and find $\mathcal{C}_{QP}=5$. (c) We choose $\nu=3/7,N_e=13,N_x=5,N_y=6,\gamma=0.5$ and find $\mathcal{C}_{QP}=7$. (d) We choose bosonic $\nu=1/2,N_e=18,N_x=7,N_y=5,\gamma=1.0$ and find $\mathcal{C}_{QP}=2$.}
    \label{BC}
\end{figure}

We evaluate $\Omega_{\rm QP}(\mathbf k)$ using the CF wave functions for the fermionic Jain states at $\nu=1/3$, $2/5$, and $3/7$, as well as for the bosonic Laughlin state at $\nu=1/2$ (in which case $p=1/2,n=1$). As shown in Fig.~\ref{BC}, we universally find $C_{\rm QP}=2pn+1$, whereas a noninteracting electron in a Landau level gives $C=1$.

\begin{figure}[tb]
    \centering    
    \includegraphics[width=0.95\linewidth]{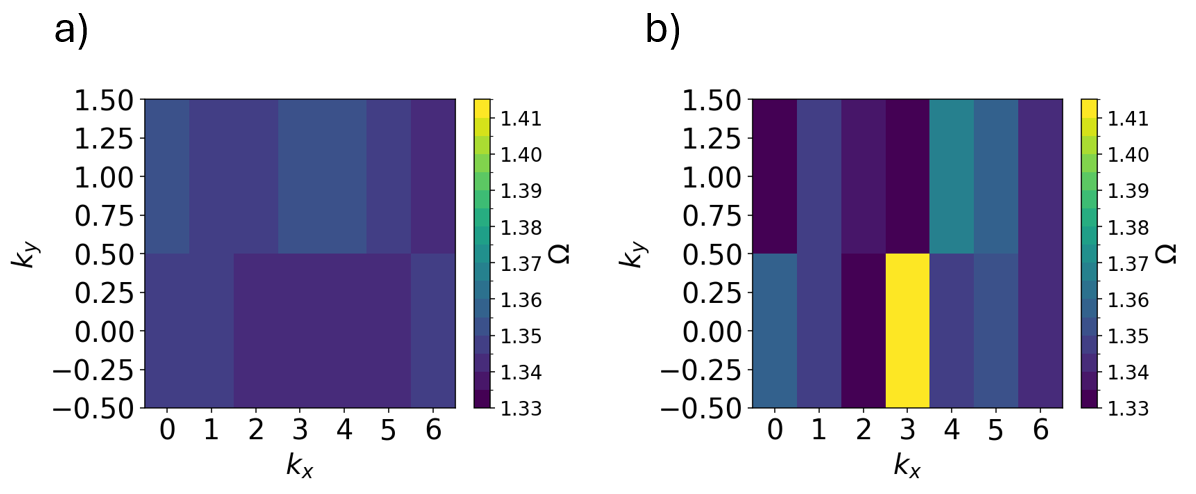}
    \caption{Plaquette Berry flux for 1 qp $N_e=5, \nu=1/3$ $\gamma=0$ (a) and $\gamma=0.2$ (b). The total Berry phases are both $6\pi$. The small variations of (a) come from Monte Carlo statistical error. }
    \label{berry flux}
\end{figure}

Furthermore, we find that the distribution of the Berry flux is uniform when $\gamma=0$ and becomes nonuniform when $|\gamma|>0$, as shown in Fig.~\ref{berry flux}. In the next section, we calculate the Berry flux in twisted bilayer MoTe$_2$, which is found to be consistent with our results.

\section{Continuum model and quasiparticle-band topology in twisted \texorpdfstring{MoTe$_2$}{MoTe2}}
\label{arxiv-sm-section-6}
\label{sec:mote2-qp}

To further test whether the $\mathcal{C}_{QP}=2pn+1$ result also holds for realistic materials, we use ED to calculate the QP band Berry flux of twisted bilayer MoTe$_2$ at $\nu=1/3$. In this section, we describe the continuum model, exact diagonalization, and calculation of the Chern number of the quasiparticle band. The continuum-model parameters and the single-band projection follow Ref.~\cite{Wang2024Fractional}, while the Berry flux is evaluated using Eq.~\eqref{EBC} with the many-body eigenstates obtained from ED.

We consider an R-stacked twisted homobilayer system. In the layer basis, the continuum Hamiltonian for one valley is
\begin{equation}
    \mathcal H_K(\mathbf r)=
    \begin{pmatrix}
        -\dfrac{\hbar^2(\hat{\mathbf k}-\boldsymbol\kappa_b)^2}
        {2m^*}+\Delta_b(\mathbf r)
        & \Delta_T(\mathbf r) \\
        \Delta_T^*(\mathbf r)
        & -\dfrac{\hbar^2(\hat{\mathbf k}-\boldsymbol\kappa_t)^2}
        {2m^*}+\Delta_t(\mathbf r)
    \end{pmatrix},
    \label{eq:mote2-continuum}
\end{equation}
where $\hat{\mathbf k}=-i\boldsymbol\nabla$ and $\boldsymbol\kappa_{b,t}$ denote the valley centers of the bottom and top layers. The intralayer moir\'e potentials and interlayer tunneling are
\begin{equation}
    \Delta_{b/t}(\mathbf r)
    =2v\sum_{j=1,3,5}\cos(\mathbf G_j\cdot\mathbf r\pm\psi),
    \qquad
    \Delta_T(\mathbf r)
    =w\left(1+e^{-i\mathbf G_2\cdot\mathbf r}
               +e^{-i\mathbf G_3\cdot\mathbf r}\right),
    \label{eq:mote2-potentials}
\end{equation}
with the upper (lower) sign corresponding to the bottom (top) layer. Here,
\begin{equation}
    \mathbf G_j=\frac{4\pi}{\sqrt{3}a_M}
    \left(\cos\frac{\pi(j-1)}{3},\,
          \sin\frac{\pi(j-1)}{3}\right),
    \qquad
    a_M=\frac{a}{2\sin(\theta/2)},
    \label{eq:mote2-reciprocal}
\end{equation}
where $a$ is the monolayer lattice constant and $\theta$ is the twist angle. We adopt the relaxed-structure parameter set of Ref.~\cite{Wang2024Fractional},
\begin{equation}
    m^*=0.6m_e,\qquad a=3.52\,\text{\AA},\qquad
    v=20.8\,\mathrm{meV},\qquad
    w=-23.8\,\mathrm{meV},\qquad \psi=107.7^\circ.
    \label{eq:mote2-parameters}
\end{equation}
The phase in Eq.~\eqref{eq:mote2-parameters} is specified in the layer convention of Eq.~\eqref{eq:mote2-potentials}. The opposite valley is related by time reversal.

\begin{figure}[tb]
    \centering
    \includegraphics[width=0.95\linewidth]{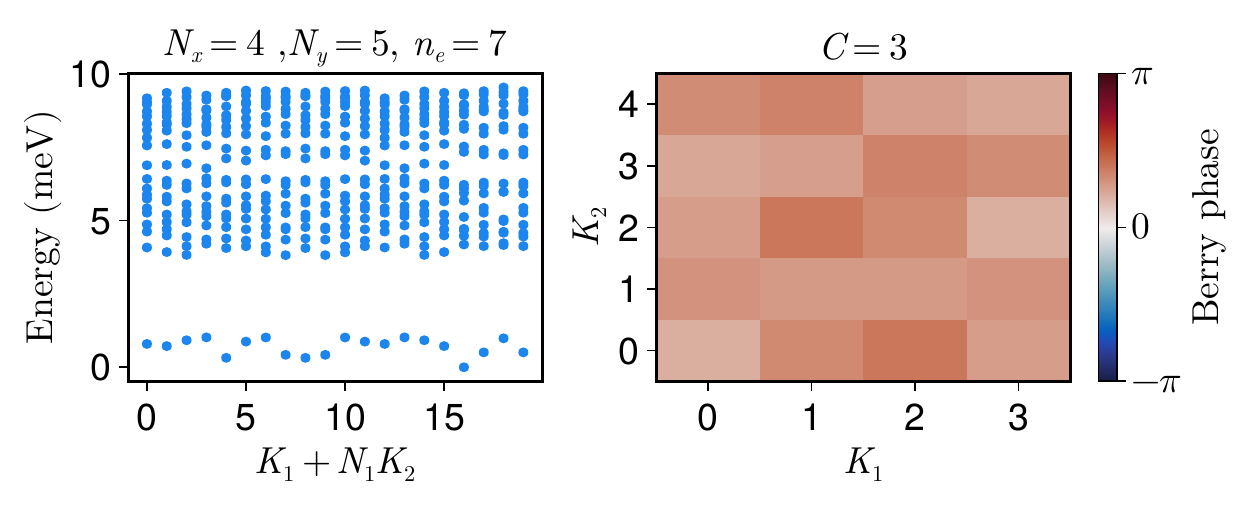}
    \caption{Quasiparticle-band dispersion and topology in twisted bilayer MoTe$_2$ for $(N_x,N_y)=(4,5)$ and $N_e=7$ holes, with $\theta=3.5^\circ$ and $\epsilon=10$. Left: The lowest 20 many-body energies in each momentum sector, plotted against $K_1+N_xK_2$. Energies are measured relative to the minimum energy of the QP band. The lowest state in each sector forms an isolated dispersive QP band. Right: Berry flux $\Omega_{\mathrm{QP}}(\mathbf K)$ through each momentum-space plaquette, evaluated using Eq.~\eqref{EBC}. Summing over all 20 plaquettes gives a total flux of $6\pi$ and a QP-band Chern number $\mathcal C_{\mathrm{QP}}=3$.}
    \label{fig:mote2-qp-chern}
\end{figure}

\begin{figure}[tb]
    \centering
    \includegraphics[width=0.95\linewidth]{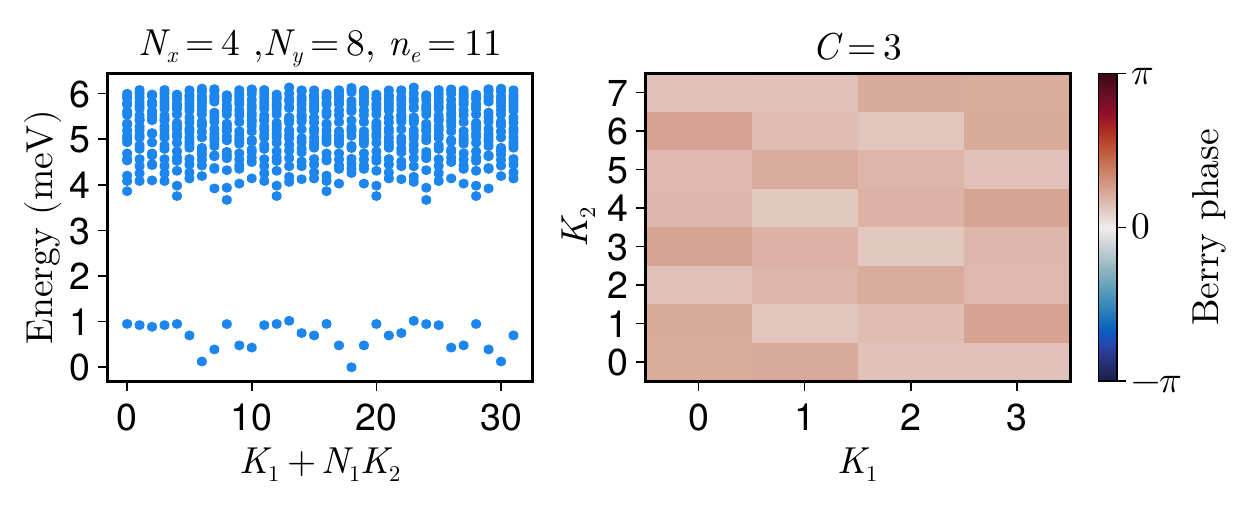}
    \caption{Quasiparticle-band dispersion and topology in twisted bilayer MoTe$_2$ for $(N_x,N_y)=(4,8)$ and $N_e=11$ holes, with $\theta=3.5^\circ$ and $\epsilon=10$. Left: The lowest 20 many-body energies in each momentum sector, plotted against $K_1+N_xK_2$. Energies are measured relative to the minimum energy of the QP band. The lowest state in each sector forms an isolated dispersive QP band. Right: Berry flux $\Omega_{\mathrm{QP}}(\mathbf K)$ through each momentum-space plaquette, evaluated using Eq.~\eqref{EBC}. Summing over all 32 plaquettes gives a total flux of $6\pi$ and a QP-band Chern number $\mathcal C_{\mathrm{QP}}=3$.}
    \label{fig:mote2-qp-chern1}
\end{figure}

We diagonalize Eq.~\eqref{eq:mote2-continuum} in a plane-wave basis and project the interacting problem onto its topmost valence band. Throughout this section, $N_e$ denotes the number of active carriers, which are holes in MoTe$_2$. In the hole-energy convention used in the numerical implementation, the one-body dispersion is $\varepsilon_{\mathbf k}=-E_v(\mathbf k)$, up to an overall energy reference, where $E_v(\mathbf k)$ is the topmost valence-band energy. The projected Hamiltonian is
\begin{equation}
    H_{\mathrm{proj}}
    =\sum_{\mathbf k}\varepsilon_{\mathbf k}
        c_{\mathbf k}^{\dagger}c_{\mathbf k}
    +\frac{1}{2A}\sum_{\mathbf q\ne0}\mathcal V(\mathbf q)
        :\rho(\mathbf q)\rho(-\mathbf q):\,,
    \qquad
    \mathcal V(\mathbf q)=\frac{e^2}{2\epsilon_0\epsilon q}\tanh{(qd)},
    \label{eq:mote2-projected}
\end{equation}
where $c_{\mathbf k}^{\dagger}$ creates an active carrier in the retained band, $A=N_xN_y A_M$, $A_M=\sqrt{3}a_M^2/2$ is the moir\'e unit-cell area, and the colons denote normal ordering in the active-carrier basis. We use a relative dielectric constant $\epsilon=10$ and $d=30$ nm. The projected density operator and its form factor are
\begin{equation}
    \rho(\mathbf q)
    =\sum_{\mathbf k}F(\mathbf k,\mathbf q)
      c_{\mathbf k+\mathbf q}^{\dagger}c_{\mathbf k},
    \qquad
    F(\mathbf k,\mathbf q)
    =\langle\psi_{\mathbf k+\mathbf q}|
        e^{i\mathbf q\cdot\mathbf r}|\psi_{\mathbf k}\rangle,
    \label{eq:mote2-density}
\end{equation}
where $|\psi_{\mathbf k}\rangle$ is the normalized Bloch orbital used for the band projection, and $\mathbf k+\mathbf q$ is understood to be reduced into the chosen moir\'e Brillouin zone. We impose periodic boundary conditions on an $N_x\times N_y$ moir\'e supercell and block diagonalize $H_{\mathrm{proj}}$ using particle number and the two total crystal-momentum indices, $K_1=\sum_i m_{1,i}\pmod{N_x}$ and $K_2=\sum_i m_{2,i}\pmod{N_y}$, where $(m_1,m_2)$ label the single-particle momentum mesh. We assume complete spin-valley polarization and retain only one valley. A single QP of the parent state at one-third hole filling is obtained by choosing
\begin{equation}
    N_xN_y=3N_e-1.
    \label{eq:mote2-qp-count}
\end{equation}
For the result shown in Fig.~\ref{fig:mote2-qp-chern} and Fig.~\ref{fig:mote2-qp-chern1}, we use $(N_x,N_y)=(4,5),\ N_e=7$ and $(N_x,N_y)=(4,8),\ N_e=11$. This branch is required to have a nonzero direct gap to the next state at every $\mathbf k$. As shown in Fig.~\ref{fig:mote2-qp-chern} and Fig.~\ref{fig:mote2-qp-chern1}, the lowest state in each momentum sector forms an isolated dispersive band. Using Eq.~\eqref{EBC}, we obtain a total Berry flux of $6\pi$, giving $\mathcal C_{\mathrm{QP}}=3$ for the orientation specified in that equation. This result is consistent with the relation $\mathcal C_{\mathrm{QP}}=\mathcal D$, where $\mathcal D=3$ is the topological degeneracy of the parent one-third-filled Laughlin state. The invariant is defined over the QP crystal-momentum Brillouin zone at fixed physical boundary conditions; it is distinct from the many-body Chern number obtained by varying boundary twists to determine the Hall conductance.

\end{document}